\documentclass[12pt]{article}
\usepackage{amssymb}
\usepackage{bm}
\usepackage{a4}
\usepackage{graphicx}
\evensidemargin \oddsidemargin
\def\1{{\bf 1}}
\def\0{{\bf 0}}

\def \E {{\cal E}}

\def \B {{\cal B}}
\def \F {{\cal F}}
\def \xim {\xi_{{\scriptscriptstyle -}}}

\def\nn{\nonumber \\}

\newcommand{\bW}{{\bm W}}
\newcommand{\by}{{\bm y}}
\newcommand{\bY}{{\bm Y}}
\newcommand{\bx}{{\bm x}}
\newcommand{\bX}{{\bm X}}
\newcommand{\bu}{{\bm u}}

\newcommand{\bv}{{\bm v}}

\newcommand{\Ba}{{\bm \alpha}}
\newcommand{\bb}{{\bm \beta}}

\newcommand{\tbb}{\widetilde{\bm \beta}}
\newcommand{\tbv}{\widetilde{\bm v}}
\newcommand{\bj}{{\bm j}}
\newcommand{\Be}{{\bm \epsilon}}
\newcommand{\bE}{{\bm E}}
\newcommand{\bB}{{\bm B}}
\newcommand{\bA}{{\bm A}}

\newcommand{\bee}{{\bm e}}
\newcommand{\Bp}{{\bm p}}

\newcommand{\be}{\begin{equation}}
\newcommand{\ee}{\end{equation}}
\newcommand{\bea}{\begin{eqnarray}}
\newcommand{\eea}{\end{eqnarray}}
\newcommand{\ba}{\begin{array}}
\newcommand{\ea}{\end{array}}
\newtheorem{prop}{Proposition}

\def\sq{\mbox{\rlap{$\sqcap$}$\sqcup$}}
\newenvironment{proof}[1]{\vspace{5pt}\noindent{\bf Proof #1}\hspace{6pt}}%
{\hfill\sq}
\newcommand{\bp}{\begin{proof}}
\newcommand{\ep}{\end{proof}\par\vspace{10pt}\noindent}
\begin{document}

\title{On plane-wave relativistic electrodynamics in plasmas and in vacuum}

\author{   Gaetano Fiore  \\\\
Dip. di Matematica e Applicazioni, Universit\`a ``Federico II''\\
   V. Claudio 21, 80125 Napoli, Italy;\\         
 I.N.F.N., Sez. di Napoli,
        Complesso MSA, V. Cintia, 80126 Napoli, Italy}
\date{}

\maketitle

\begin{abstract}
\noindent

We revisit the exact microscopic equations (in differential, and equivalent integral form)
ruling a relativistic cold plasma after the plane-wave Ansatz,
without  customary approximations. We show 
that in the Eulerian description the motion of a very diluted plasma
initially at rest and excited by an arbitrary transverse plane electromagnetic 
travelling-wave has a very simple and explicit
dependence on the transverse electromagnetic potential;
for a non-zero density plasma the above motion is a good approximation of the real
one as long as the back-reaction of the charges on the electromagnetic field 
can be neglected, i.e. for a time lapse decreasing with the plasma density, 
and can be used as initial step in an iterative resolution scheme.
  

As one of many possible applications, we use these results to describe how the ponderomotive force of
a very intense and short plane laser pulse hitting normally the surface of a plasma boosts 
the surface electrons into the ion background.  Because of this penetration 
the electrons are then pulled back by the electric force exerted by the ions
and may leave  the plasma with high energy in the direction opposite to that of propagation of
the pulse [G. Fiore, R. Fedele, U. De Angelis, 
{\it The slingshot effect: a possible new laser-driven high energy acceleration mechanism for electrons}, 
arXiv:1309.1400].

\end{abstract}

\section{Introduction}

Consider a charged test particle initially at rest.
The solution $\bx(t,\bX)$ of its equation of motion 
under the action of a given electromagnetic (EM) field is a function of the time $t$ and of 
the initial position $\bX$. By definition
a charged test particle  has such a small electric charge that its
influence on the EM field can be neglected. This is of course an
idealization of real particles, but actually an extremely good one for microscopic particles
(electrons, ions, etc) under the action of macroscopic EM fields.
If we imagine to fill at the initial time  a  region $\B$
with a ``very diluted'' fluid of test particles of the same kind at rest, then the function  $\bx(t,\bX)$
will describe the motion of such a fluid under the action of the given EM field.
$\bX\!\in\!\B$ will be used as a label to distinguish the single particles  composing 
the fluid (Lagrangian description): $\bx(t,\bX)$ [more generally, a function $\tilde f(t,\bX)$)]
will give the position (more generally, the fluid observable $f$) at time $t$ of the 
particle (or better, fluid element) initially located at $\bX$.
By definition, the fluid just described will have  zero density and be initially at rest;
this motivates us in the first place to study its equation of motion
in a given (necessarily free) EM field. 
If the free EM field is an arbitrary transverse plane travelling-wave with a
(whether smooth or sharp) front,
then the solution has a very simple 
expression (subsection \ref{0densitysol}) in terms of the EM field 
(it is completely explicit apart from the inversion 
of a strictly increasing function $\Xi(\xi)$ of one variable $\xi$). 
We prove this result going from the Lagrangian to the Eulerian  description, and back.
The motion of a single test particle with non-zero initial
velocity is obtained from that with zero initial velocity 
by a suitable Poincar\'e transformation.
These results agree with the motion (in implicit form) of the particle
obtained by resolution of its Hamilton-Jacobi equation  \cite{LanLif62}
(see e.g. also \cite{EbeSle68,EbeSle69}).


One can combine several such fluids  into a cold plasma, and ask 
whether and how well these zero-density solutions 
approximate the solutions of the (coupled) Lorentz-Maxwell and continuity
equations (subsection \ref{general}) for non-zero densities with the same plane symmetry
and asymptotic conditions (subsection \ref{plane}). In section \ref{integral} in full
generality we solve
for the longitudinal electric field in terms of the longitudinal coordinates of the particles
and formulate   integral equations  equivalent to the remaining differential equations. 
In \cite{Fioetal} we will discuss a recursive resolution scheme for these  integral equations
in which the zero-density solutions play the role of lowest order approximations. 
In section \ref{constdensity} we just determine the first correction for a EM wave 
hitting a step-shaped density plasma and show for how long this correction
is small.  As an illustration,  in section \ref{Sling} we use
these results  to determine conditions under which
the {\it slingshot effect}  \cite{FioFedDeA13} may occur.

Up to section \ref{integral} no role is played by
Fourier analysis and related methods/notions, such as the 
Slowly Varying Amplitude Approximation \cite{Kar74,She84,Whi91}, (frequency-dependent)
refractive indices \cite{Hec02,AkhEtAl75}, etc. 
Our scope here is to derive from the basic microscopic equations
some general results which can be used, at least for a limited time lapse,
also in the presence of very intense EM waves
with completely arbitrary profile (we do not assume existence of a carrier frequency 
nor of a slowly varying envelope amplitude), when the  methods mentioned above become inadequate.
These conditions, which involve fast and highly nonlinear effects, characterize
important plasma phenomena in current laboratory research on
laser-plasma interactions,  such as Laser Wake Field 
excitations and acceleration
\cite{TajDaw79,Gorbunov-Kirsanov1987,Sprangle1988} (with external 
\cite{Irman2007} or self injection \cite{Joshi2006}; 
for a short list of their applications
see also the introduction of\cite{Fio14}), 
especially in the \textit{bubble} (or \textit{blowout}) \textit{regime} 
\cite{FauEtAl04,MalEtAl05,Kalmykov2009}, or the propagation of intense laser beams 
through overdense plasmas \cite{AkhPol56,KawDaw70}, etc.
(see e.g.  \cite{MouUms92}).
Only in section \ref{constdensity} we specialize calculations
to a very intense laser pulse with a (not so slowly) varying envelope.

\section{Preliminaries}
\subsection{Cold plasma equations}
\label{general}

We first recall the basic plasma equations and some basic 
facts about fluid mechanics, while fixing the notation.
We consider a collisionless plasma composed by
$k\ge 2$ types of charged particles (electrons, ions of various kinds, etc).
For $h\!=\!1,...,k$ we denote by $m_h,q_h$ the rest mass
and charge of the $h$-th type of particle, by $\bv_h(\bx,ct)$, $n_h(\bx,ct)$
respectively the 3-velocity and the density (number of particles -
treated as usual as a continuous rather than an integer-valued variable - per unit volume)
of the corresponding fluid element located in position $\bx$ at time $t$
($c$ is the light velocity). It is convenient to formulate
the equations in terms of dimensionless unknowns like \ $\bb_h\!:=\!\bv_h/c$,
$\gamma_h\!:=\!1/\sqrt{1\!-\!\bb_h^2}$, \
 the 4-velocity of the $h$-th type of particle \
 $( u_h^\mu)\!:=\!(\gamma_h,\gamma_h \bb_h)$. \ The latter is normalized, \
$ u_h^\mu u_{h\mu}\!=\!1$ \ (indices are raised and lowered by the
Minkowski metric $\eta_{\mu\nu}\!=\!\eta^{\mu\nu}$, with
$\eta^{00}\!=\!1$, $\eta^{11}\!=\!-1$, etc.), implying $\gamma_h\!=\!u_h^0\!=\!\sqrt{1\!+\!\bu_h^2}$,
$\bb_h\!=\!\bu_h/\gamma_h$;
$u_h^\mu$ is essentially the 4-momentum $(p^\mu_h)=(p_h^0,\Bp_h)$ of the $h$-th type of particles
made dimensionless by dividing by the appropriate powers of $m_h, c$: \
$ \gamma_h\!=\!u_h^0\!=\!p^0_h/m_hc^2$, \
$\bu_h\!=\!\mbox{\boldmath $p$}_h/m_hc$. \
The 4-vector current density \
$(j^\mu)=(j^0,\bj)$ is now given by $j^0\!=\!\sum_{h=1}^k q_h n_h$, \
$\bj\!=\!\sum_{h=1}^k q_h n_h\bv_h/c\!=\!\sum_{h=1}^k q_h n_h\bb_h$. 

We use the CGS system.
As usual, we denote by $(A^\mu)=(A^0,\bA)$ the electromagnetic potential,
by $F^{\mu\nu}=\partial^\mu
A^\nu-\partial^\nu A^\mu$ the electromagnetic field,
where $x=(x^\mu)=(x^0,\bx)=(ct,\bx)$,
$(\partial_\mu)\equiv(\partial/\partial x^\mu)=(\partial_0,\nabla)$.
We start from the explicitly Lorentz-covariant formulation of the
Maxwell's equations and  of the Lorentz equations of motion
of the collisionless fluids:
\bea
&&\Box A^\nu-\partial^\nu(\partial_\mu A^\mu)
=\partial_\mu F^{\mu\nu}=4\pi j^\nu,\label{Maxwell}\\[8pt]
&&-q_h u_{h\mu} F^{\mu\nu}=m_hc^2 u_{h\mu} \partial^\mu u_h^{\nu}
 \qquad h=1,...,k;                  \label{hom'}
\eea
the unknowns are $A^\mu,n_h, \bu_h$, \ $h=1,...,k$.
[As known, the component $\nu=0$ of  (\ref{hom'})
is not independent of the other ones;
it can be obtained contracting
the  components $\nu=l=1,2,3$ with $ u^l_h$
and using the definitions above]. Dividing the  components
$\nu=l$ by $\gamma_h$ and using the definitions 
$E^l=F^{l0}=-\partial_0A^l-\partial_l A^0$, $B^l=-\frac 12
\varepsilon^{lkn}F^{kn}=\varepsilon^{lkn}\partial_k A^n$ 
of the electric and magnetic field one obtains
the equivalent, familiar 3-vector formulation of (\ref{hom'})
\be
q_h\left(\bE +\frac{\bv_h}c \wedge \bB
\right)=\partial_t\Bp_h+\bv_h \cdot \nabla\Bp_h
=\frac {d\Bp_h}{dt}, \qquad\qquad
\Bp_h=m_h\gamma_h \bv_h.
\label{hom}
\ee
In the second equality  we have used the relation between
the {\it total} (or {\it material}) derivative $\frac d{dt}$
for the $h$-th type of particle and $\partial_\mu$, $\bv_h$:
\be
\frac {d}{dt}:=\partial_t\!+\! v^l_h\partial_l
\qquad\qquad \Leftrightarrow \qquad \qquad \frac {d}{dx^0}:=\partial_0\!+\!
\beta^l_h\partial_l=\frac{u_h^\mu}{\gamma_h}\partial_\mu.             \label{totder}
\ee
Eq. (\ref{hom'}) are formulated in the Eulerian description: the volume element  of
the $h$-th fluid is labelled by its position $\bx$ at time $x^0=c t$,
and all the observables are described as functions of $(x^0,\bx)$.
Given an observable $f_h(x^0,\bx)$ of the $h$-th fluid in
the Eulerian description, it is related to its Lagrangian counterpart $\tilde f_h(x^0,\bX)$ by
\be
\tilde f_h(x^0,\bX)=f_h[x^0,\bx_h(x^0,\bX)]\qquad\Leftrightarrow\qquad
f_h(x^0,\bx)=\tilde f_h[x^0,\bX_h(\bx,x^0)];
\ee
here $\bx_h(x^0,\bX)$ is the position at time  $x^0=c t$ of the $h$-th fluid element
initially located in $\bX$, and $\bX_h(x^0,\bx)$ is its inverse (for fixed $x^0$),
i.e. the initial position  of the $h$-th fluid element located in $\bx$ at time  $x^0=c t$;
by construction $\bX_h(X^0, \bx )=\bx$.
The correspondences \ $\bx_h(x^0, \cdot):\bX\mapsto \bx$ \ and \
$\bX_h(x^0, \cdot):\bx\mapsto \bX$ \ are required to be one-to-one and to have
continuous first derivatives (so the Jacobian
$\tilde J_h:=\det\Vert\frac {\partial \bx_h} {\partial \bX}\Vert$
as well as its inverse never vanish and reduce to 1 at $x^0= X^0 $) 
and continuous second derivatives with respect to time (at least piecewise). Clearly,
$J_h=\det^{-1}\Vert\frac {\partial \bX_h}{\partial \bx} \Vert$.
By construction, $f_h(X^0, \bx )\!=\!
\tilde f_h(X^0, \bx )\!=:\!\tilde f_{h0}(\bx)$, where $X^0$ is the initial time.
In the Lagrangian and Eulerian description
the conservation of the number of particles of the $h$-th fluid in each material
volume element of volume $d^3x$  respectively amounts to
\be
\tilde n_h(x^0,\bX) \tilde J_h(x^0,\bX)= \widetilde{n_{h0}}(\bX)\qquad\qquad
\Leftrightarrow \qquad \qquad n_h(x^0,\bx) J_h(x^0,\bx)_{h0}(x^0,\bx). \label{n_hg}
\ee
$\widetilde{n_{h0}}$  is $\widetilde{n_h}$ 
evaluated  at the time $x^0=X^0$; it is part of the initial data. 
As known, $\frac d{dt}$ becomes the partial derivative $\partial_t$
(or equivalently, $\frac d{dx^0}$ becomes $\partial_0$)
when going from the Eulerian to the Lagrangian description.
The continuity equation  of the $h$-th fluid follows from (\ref{n_hg}) and
$\partial_0 \widetilde{n_{h0}}=0$, and reads
\be
\partial_0 (\tilde n_h \tilde J_h)=0,
\qquad\qquad\Leftrightarrow \qquad \qquad
\frac {d n_h}{dx^0}\!+\!n_h\nabla\!\cdot\!\bb_h
=\partial_0n_h\!+\!\nabla\!\cdot\!(n_h\bb_h)=0.    \label{clh'}
\ee
The equivalence in (\ref{clh'}) is based on the well-known identity
$\frac {d J_h}{dx^0}=J_h \nabla\!\cdot\!\bb_h.       
$
Eq. (\ref{Maxwell}) implies the electric current conservation law
$\partial_\mu j^\mu=0$, what makes one of
the $k$ conservation laws (\ref{clh'}) redundant.
From the definition
$\tbv_h(ct,\bX):=\partial_t \bx_h( ct,\bX)$, or equivalently
$\tbb_h(x^0,\bX):=\partial_0 \bx_h(x^0,\bX)$, it follows that
if the function $\bb_h(x^0,\bx)$ is known one can determine
the maps $\bx_h:\bX\mapsto \bx$
by solving the Cauchy problems
\bea
\partial_0 \bx_h(x^0,\bX)=\bb_h\!\left[{x^0},\bx_h({x^0},\bX) \right]  \label{lageul}
\eea
with initial conditions $\bx_h(X^0,\bX)\!=\!\bX $. 
If all the observables $f_h$ admit for all $\bx$ finite limits 
$\lim_{x^0\to-\infty}f_h(x^0,\bx)\!=:\!\tilde f_{h0}(\bx)$ and in particular
$\tilde \bb_{h0}(\bx)\!\equiv\!\0$,
then one may use also $X^0\!=\!-\infty$ as `initial' time,
and as `initial' conditions the corresponding asymptotic ones, e.g. for
each $\bX$ the condition
$\bx_h(x^0,\bX)\!\to\!\bX$ as $x^0\!\to\!-\infty$ to complement (\ref{lageul}).

\subsection{Cold plasma equations for plane waves with a front}
\label{plane}

Henceforth 
 we restrict our attention to solutions of (\ref{Maxwell}-\ref{hom}),  (\ref{n_hg}-\ref{lageul})
such that for all $h$:
\bea
&& A^\mu, n_h, \bu_h \qquad \mbox{{\bf  depend only on }} z\!\equiv\!
x^3\!,x^0 \qquad \mbox{(plane wave Ansatz)},     \label{pw}\\[10pt]
&&\!\!\ba{ll}
A^\mu(x^0\!,z)\!=\!0,\qquad \bu_h(x^0\!,z)\!=\!\0,&\qquad\qquad \mbox{if }\:\:x^0\!\le\! z,\\ [4pt]
\exists\:\:\widetilde{ n_{h0}}(z) \qquad \mbox{such that }
\:\displaystyle\sum_{h=1}^k\!q_h\widetilde{ n_{h0}}
\!\equiv\!0, \qquad n_h(x^0\!\!,z)\!=\!\widetilde{ n_{h0}}(z)&\qquad\qquad \mbox{if }\:\:x^0\!\le\! z.
\ea
  \qquad         \label{asyc}
\eea
These equations (which entail a partial gauge-fixing\footnote{The class of $A^\mu$ 
depending only on $z,x^0$ is invariant only under  gauge transformations $A^\mu\!\to\!
A^\mu\!+\!\partial^\mu \Lambda$ with $\Lambda$ depending only on
$z,x^0$. No further gauge-fixing is done in this section, so our
results are invariant under the latter transformations. Among the
possible choices in the class there is the Coulomb gauge, which
satisfies in addition the condition $\partial_zA^z=0$.})  imply
\bea
&&\bB\!=\!\bB^{{\scriptscriptstyle\perp}}\!=\!{\hat{\bm z}}\!\wedge\!\partial_z\bA\!^{{\scriptscriptstyle\perp}},
\qquad\qquad\qquad\bE^{{\scriptscriptstyle\perp}}\!=\!-\partial_0\bA\!^{{\scriptscriptstyle\perp}},\qquad  \label{conseq}\\[8pt]
&& \bE(x^0\!,z)=\bB(x^0\!,z)=\0\qquad\qquad
\bx_h(x^0\!,\bx)=\bx \qquad\qquad\qquad \mbox{if }\:\:x^0\le z ,\qquad
 \label{conseq'}
\eea
and  are fulfilled, after  a suitable $z$-translation, by any plasma 
initially neutral and at equilibrium with purely $z$-dependent densities $\widetilde{ n_{h0}}(z)$
in a zero electromagnetic field and then reached by
a purely transverse electromagnetic wave (with a front) propagating from infinity 
in the positive $z$ direction.
Of course, our results will be applicable also if conditions (\ref{pw}-\ref{asyc}) can be achieved 
by a Poincar\'e transformation.
Equations (\ref{asyc})$_3$,  (\ref{conseq'})$_2$ trivially  imply \
$n_h(-\infty,Z)\!=\!\widetilde{ n_{h0}}(Z)$, \
$\bx_h(-\infty,\bX)\!=\!\bX$. \  Therefore we can adopt also
the configuration of the plasma at the  `initial' time $X^0\!=\!-\infty$ as a 
reference configuration in the Lagrangian description.

$\bA\!^{{\scriptscriptstyle\perp}}$ has become {\bf  a physical observable}, as it can be recovered
from  $\bE^{{\scriptscriptstyle\perp}}$ integrating (\ref{conseq})$_1$ with respect to $x^0$ at fixed $z$ and
exploiting (\ref{asyc}): $-\bA\!^{{\scriptscriptstyle\perp}}(x^0,z)=\!\int^{x^0}_{z}\!\!\!\!dx^0{}' \,
\bE^{{\scriptscriptstyle\perp}}(x^0{}',z)=\!\int^{x^0}_{-\infty }\!\!\!\!dx^0{}' \,
\bE^{{\scriptscriptstyle\perp}}(x^0{}',z)$.

Eq. (\ref{hom'})$_{\nu=x,y}$ implies
\vskip-.3cm
$$
\qquad\qquad\qquad u_h^\mu \partial_\mu
\!\left(m_hc^2\bu_h^{{\scriptscriptstyle\perp}}
\!+\!q_h\bA\!^{{\scriptscriptstyle\perp}}\!\right)\!=\!\gamma_h\frac {d(m_hc^2
\bu_h^{{\scriptscriptstyle\perp}}\!+\!q_h\bA\!^{{\scriptscriptstyle\perp}})}{dx^0}\!=\!0;
$$
in the Lagrangian description this becomes $\partial_0(m_hc^2
\tilde{\bu}_h^{{\scriptscriptstyle\perp}}\!+\!q_h\widetilde{\bA}\!^{{\scriptscriptstyle\perp}})=0$, implying that
$m_hc^2\tilde{\bu}_h^{{\scriptscriptstyle\perp}}\!+\!
q_h\tilde{\bA}\!^{{\scriptscriptstyle\perp}}
=C(\bX)=$const \ with respect to $x^0$; \ by  (\ref{asyc})  \  $C(\bX)\equiv 0$.
Thus one obtains the known result
\be
\tilde{\bu}_h^{{\scriptscriptstyle\perp}}\!= \frac {-q_h}{m_hc^2}\tilde{\bA}\!^{{\scriptscriptstyle\perp}}\qquad
\Leftrightarrow\qquad\bu_h^{{\scriptscriptstyle\perp}}\!= \frac {-q_h}{m_hc^2}\bA\!^{{\scriptscriptstyle\perp}},
 \label{hom'12}
\ee
which explicitly gives
$\bu_h^{{\scriptscriptstyle\perp}}$ in terms of $\bA\!^{{\scriptscriptstyle\perp}}$.
Eq. (\ref{Maxwell}) and the remaining (\ref{hom'}) become
\bea
&& (\ref{Maxwell})_{\nu=0}:\qquad\partial_z E^z=4\pi \sum_{h=1}^kq_hn_h
,\label{Maxwell0}\\[8pt]
&& (\ref{Maxwell})_{\nu=z}:\qquad \partial_0E^z=-\frac{4\pi}c
\sum_{h=1}^kq_hn_hv^z_h=-4\pi \sum_{h=1}^kq_hn_h\beta^z_h,
\label{Maxwell3}\\[8pt]
&& (\ref{Maxwell})_{\nu=x,y}:\quad\:
\left(\partial_0^2\!-\!\partial_z^2\right)\!\bA\!^{{\scriptscriptstyle\perp}}\!\!-\! 4\pi
\sum_{h=1}^kq_hn_h\bb^{{\scriptscriptstyle\perp}}_h\stackrel{(\ref{hom'12})}{=}\!
\left[\partial_0^2\!-\!\partial_z^2\right]\!\!\bA\!^{{\scriptscriptstyle\perp}}
\!+\! \frac{4\pi }{c^2}\left[\sum_{h=1}^k
\frac{q_h^2n_h}{m_h\gamma_h}\right]
\!\bA\!^{{\scriptscriptstyle\perp}}\!=0,\qquad\qquad \label{Maxwell12}\\[8pt]
&& (\ref{hom'})_{\nu=0}:\qquad  m_hc^2u_h^\mu \partial_\mu \gamma_h\!-\!q_hu^z_hE^z
 \!+\!q_h \bu_h^{{\scriptscriptstyle\perp}}\cdot\partial_0\bA\!^{{\scriptscriptstyle\perp}}
\stackrel{(\ref{totder}),(\ref{hom'12})}{=}
 m_hc^2\gamma_h\frac{d\gamma_h}{dx^0} \!-\!q_hu^z_hE^z \nn[8pt]&& \qquad\qquad\qquad
-\frac{q_h^2}{m_hc^2} \bA\!^{{\scriptscriptstyle\perp}}\!\cdot\!\partial_0\bA\!^{{\scriptscriptstyle\perp}}=0,
 \qquad \Leftrightarrow\qquad
\frac{d\gamma_h}{dx^0}-\frac{q_hu^z_h E^z}{m_hc^2\gamma_h}-
\frac{q_h^2\partial_0(\bA\!^{{\scriptscriptstyle\perp}})^2}{2\gamma_hm_h^2c^4}=0\label{hom'0}\\[8pt]
&& (\ref{hom'})_{\nu=z}:\qquad  m_hc^2u_h^\mu \partial_\mu u^z_h\!-\!q_h\gamma_hE^z
 \!-\!q_h \bu_h^{{\scriptscriptstyle\perp}}\cdot\partial_z\bA\!^{{\scriptscriptstyle\perp}}
\stackrel{(\ref{totder}),(\ref{hom'12})}{=}
 \gamma_h\!\left(\!m_hc^2\frac{du^z_h}{dx^0} \!-\!q_hE^z\!\right)
\nn[8pt]&& \qquad\qquad\qquad
+\frac{q_h^2}{m_hc^2}
\bA\!^{{\scriptscriptstyle\perp}}\!\cdot\!\partial_z\bA\!^{{\scriptscriptstyle\perp}}=0,\qquad \Leftrightarrow\qquad
\frac{du^z_h}{dx^0}-\frac{q_h E^z}{m_hc^2}+
\frac{q_h^2\partial_z(\bA\!^{{\scriptscriptstyle\perp}})^2}{2\gamma_hm_h^2c^4}=0\label{hom'3}
\eea
The independent unknowns in (\ref{Maxwell0}-\ref{hom'3}) are 
$\bA\!^{{\scriptscriptstyle\perp}}, u_h^z,E^z$, all observables.
We neither need nor care to determine  $A^0,A^z$ such that
$ E^z\!=\!-\partial_0A^z\!-\!\partial_z A^0$
by completing the gauge-fixing.

The invertibility of  $\bx_h:\bX\mapsto \bx$ for each fixed $x^0$ amounts
to $z_h( x^0\!,Z)$ being strictly increasing  with respect to
$ Z\equiv X^3$ for each fixed $x^0$.
We shall abbreviate $Z_h\equiv X^3_h$.
Eq. (\ref{n_hg}) 
becomes
\bea
&& \tilde n_h (x^0,Z)\, \partial_Z z_h( x^0,Z) \, =\, \widetilde{n_{h0}}( Z ),
\qquad\qquad \Leftrightarrow \qquad\qquad
n_h\, = \, n_{h0}\, \partial_z  Z_h.    \label{n_h}
\eea
$\partial_0  Z =0$ in the Eulerian description becomes \ $\frac {d  Z_h}{dx^0}= \partial_0Z_h+\beta_h^z \partial_z  Z_h=0$, \ which by  (\ref{n_h}) implies \
\be
n_{h0} \,\partial_0  Z_h\!+\!n_h\beta^z_h=0.                        \label{j_h}
\ee

\subsection{The zero-density solutions}
\label{0densitysol}

Here we determine plane wave solutions with all  $n_h\!\equiv\! 0$:  
the electromagnetic field  affects the evolution of the $\bu_h$, while the plasma
does not affect  the evolution of the field.

Clearly, $n_h\!\equiv\! 0$ for all $h$ fulfills (\ref{n_h})  and implies
$j^\mu\!\equiv\! 0$, whereby equations (\ref{Maxwell0}-\ref{Maxwell12}) become
\be
\partial_zE^z=0,\qquad\qquad\partial_0E^z=0,\qquad\qquad
\left(\partial_0^2-\partial_z^2\right)\bA ^{{\scriptscriptstyle\perp}}=0        \label{n=0}
\ee
(Maxwell equations in vacuum). The general solution fulfilling  (\ref{asyc})
has the form
\be
 E^z\equiv 0, \qquad\qquad
\bA\!^{{\scriptscriptstyle\perp}}(x^0,z)=\Ba\!^{{\scriptscriptstyle\perp}}(x^0\!-\!z),   \label{solvn=0}
\ee
with some function $\Ba\!^{{\scriptscriptstyle\perp}}\!(\xi)$ such that \
 $\Ba\!^{{\scriptscriptstyle\perp}}\!(\xi)\!=\! 0$ if $\xi\!\le\! 0$; \ this implies $\bE^{{\scriptscriptstyle\perp}}(x^0\!,z)=\Be\!^{{\scriptscriptstyle\perp}}\!(x^0\!-\!z)$, with 
$\Be^{{\scriptscriptstyle\perp}}\!(\xi)\!:=\!-\Ba\!^{{\scriptscriptstyle\perp\prime}}(\xi)$.  
Viceversa, given  $\Be\!^{{\scriptscriptstyle\perp}}\!(\xi)$ such that \
 $\Be\!^{{\scriptscriptstyle\perp}}\!(\xi)\!=\! 0$ if $\xi\!\le\! 0$, it is $\Ba^{{\scriptscriptstyle \perp}}\!(\xi)\!=\!-\!\!\int^{\xi}_{-\infty }\!\!d\xi' \Be^{{\scriptscriptstyle\perp}}\!(\xi')$.

Now we prove that for any 
$\Ba^{{\scriptscriptstyle \perp}}(\xi)\in C^2(\mathbb{R},\mathbb{R}^2)$ such that
$\Ba^{{\scriptscriptstyle \perp}}(\xi)=0$ if \ $\xi\le 0$ \ the functions
\be
\ba{lll}
\bA\!^{{\scriptscriptstyle\perp}}(x) \!=\! 
\Ba^{{\scriptscriptstyle \perp}}(\xi),\quad\:\: \xi\!\!:=\!x^0\!\!-\!z,\qquad &
n_h \!\equiv\!0,\qquad \: &
E^z \!=\!E^{z{\scriptscriptstyle (0)}}\!\!:=\! 0,   \\[8pt]
\bu_h^{{\scriptscriptstyle\perp}}(x)\!=\!\bu_{h}^{{\scriptscriptstyle \perp(0)}}(\xi)\!:=\!
\frac {-q_h}{m_hc^2}\Ba^{{\scriptscriptstyle \perp}}(\xi),
\quad \: &
u_h^z\!=\!u_{h}^{{\scriptscriptstyle z(0)}}
\!\!:=\!\frac 12\bu_{h}^{{\scriptscriptstyle \perp(0)}}{}^2\!, \qquad \:
&  \gamma_h\!=\!\gamma_{h}^{{\scriptscriptstyle (0)}}
\!\!:=1+\!u_{h}^{{\scriptscriptstyle z(0)}},
 \ea \label{n=0'+}
\ee
which depend on $x$ only through $\xi$, solve
(\ref{hom'12}-\ref{hom'3}) and  (\ref{asyc}).
In fact, the difference of eqs. (\ref{hom'0}-\ref{hom'3}) gives the equivalent equation
\bea
&& \frac {d(\gamma_h- u^z_h)}{d{x^0}}=\frac {q_h^2}
{2m_h^2c^4\gamma_h} \left(\partial_0+ \partial_z\!\right)\bA\!^{{\scriptscriptstyle\perp}}{}^2
\!-\!\frac{\gamma_h\!-\!u^z_h}{\gamma_h}
\frac{q_h E^z}{m_hc^2}
\label{bla'}
\eea
[recall that (\ref{hom'}) had only 3 independent components].
Both terms at the right-hand side (rhs) vanish by (\ref{solvn=0}), i.e. by (\ref{n=0'+})$_{1,3}$, as
 $(\partial_z\!+\!\partial_0)\Ba\!^{{\scriptscriptstyle\perp}}{}^2=0$; hence,
$\frac {d(\gamma_h\!-\! u_h^z)}{d{x^0}}=0$, or, in Lagrangian notation,
$\partial_0(\tilde\gamma_h\!-\!\tilde u_h^z)=0$. The latter equation is
solved by $\tilde\gamma_h\!-\!\tilde u_h^z\!=\!C(\bX)\!=$const \ with respect to $x^0$; \
$C(\bX)\!=\!1$ \ by  (\ref{asyc}).
This gives (\ref{n=0'+})$_6$. Eq. (\ref{n=0'+})$_5$ follows from
$$
1\!+\!\bu_h^2=\gamma_h^2\stackrel{(\ref{n=0'+})_6}{=}(1\!+\!u_h^z)^2.
$$
We have already proved that (\ref{n=0'+})$_{1-3}$ solve
(\ref{Maxwell0}-\ref{Maxwell12})    and (\ref{n_h}).
By (\ref{n=0'+})$_1$,  (\ref{n=0'+})$_4$ solves  (\ref{hom'12}) .

Eq. (\ref{n=0'+}) describe travelling-waves moving in the positive
$z$ direction (forward waves)  with phase velocity equal to the
velocity of light $c$. They are 
determined solely by the propagating electromagnetic potential $\Ba^{{\scriptscriptstyle \perp}}$.

{\bf Remarks 1.} \ If $\bE^{{\scriptscriptstyle \perp}}$ is linearly polarized in the $x$-direction
all particles' motions will be parallel to the $xz$ plane; if it is approximately periodic with zero mean, the transverse
motions will approximately average to zero. On the other hand, 
 at no time any particle can move in the negative $z$-direction because $u_{h}^{{\scriptscriptstyle z(0)}},\beta_{h}^{{\scriptscriptstyle z(0)}}$ are nonnegative-definite; the latter are the result of the
acceleration by the ponderomotive force.
The direction of the 3-velocity $\bb_h$ (and $\bv_h$) of the $h$-th fluid
approaches the $xy$ plane  as
$|\Ba^{{\scriptscriptstyle \perp}}|$ or $\frac{|q_h|}{m_h}$ decreases,  and
approaches that of the $z$-axis as
$|\Ba^{{\scriptscriptstyle \perp}}|$ or  $\frac{|q_h|}{m_h}$ grows, since
\be
\beta_{h}^{{\scriptscriptstyle z(0)}}=\frac { \bu_{h}^{{\scriptscriptstyle \perp(0)}}{}^2}
{2\!+\!\bu_{h}^{{\scriptscriptstyle \perp(0)}}{}^2},
\quad \bb_{h}^{{\scriptscriptstyle\perp(0)}}=
\frac {2\bu_{h}^{{\scriptscriptstyle \perp(0)}}}{2\!+\!
\bu_{h}^{{\scriptscriptstyle \perp(0)}}{}^2}, \qquad\Rightarrow \qquad
\frac{|\bb_{h}^{{\scriptscriptstyle\perp(0)}}|}{|\beta_{h}^{{\scriptscriptstyle z(0)}}|}=
\frac {2}{|\bu_{h}^{{\scriptscriptstyle \perp(0)}}|} =\frac {2m_hc^2}
{|q_h\Ba^{{\scriptscriptstyle \perp}}|},
\label{velocity+-}
\ee

In   (\ref{n=0'+}) the assumption $\Ba^{{\scriptscriptstyle \perp}}(\xi)\in C^2(\mathbb{R},\mathbb{R}^2)$
just guarantees that $\partial_0^2\bA^{{\scriptscriptstyle \perp}}$, $\partial_z^2\bA^{{\scriptscriptstyle \perp}}$
are separately well-defined (and equal to each other).
Assuming less regularity for $\Ba^{{\scriptscriptstyle \perp}}(\xi)$ leads to
weak solutions; in particular one can consider $\Ba^{{\scriptscriptstyle \perp}}(\xi)\in C(\mathbb{R},\mathbb{R}^2)$
with $\Ba^{{\scriptscriptstyle \perp\prime}}\!=\!\Be^{{\scriptscriptstyle \perp}}$ 
defined and continuous everywhere except in a finite number of points of  finite discontinuities (e.g. at the wavefront). 
Apart from $\Ba\!^{{\scriptscriptstyle\perp}}(\xi)= 0$ if $\xi\le 0$, the profile
of $\Ba^{{\scriptscriptstyle \perp}}(\xi)$ is completely arbitrary.

\medskip
Let us introduce the following primitives of \ $\bu_h^{{\scriptscriptstyle (0)}},\gamma_h^{{\scriptscriptstyle (0)}}$:
\be
\bY\!_h(\xi)\!:=\!\int^\xi_0\!\!\!\! d\xi' \,\bu_h^{{\scriptscriptstyle (0)}}\!(\xi'),
\qquad\quad\Xi_h(\xi)\!:=\!\int^\xi_0\!\!\!\!  d\xi'\, \gamma_h^{{\scriptscriptstyle (0)}}(\xi')\!=\!
\xi  \!+\! Y^3_h(\xi).                                        \label{defYXi}
\ee
As $u_{h}^{{\scriptscriptstyle z(0)}}\!\ge\! 0$, \ $Y^3_h(\xi)$ \ is increasing, \
$\Xi_h(\xi)$ \ is strictly increasing, therefore invertible.

\begin{prop} Choosing $\bu_h\!\equiv\!\bu_h^{{\scriptscriptstyle (0)}}$,
the solution \ $\bx^{{\scriptscriptstyle (0)}}_{h}(x^0,\bX)$ \ of the ODE
(\ref{lageul}) with the  asymptotic  condition \
$\bx_h(x^0,\bX)\!=\!\bX  {}$ \ for $x^0\!\le\! Z$,
and (for fixed $x^0$) its inverse \ $\bX^{{\scriptscriptstyle (0)}}_{h}(x^0,\bx)$ \ are given by:
\be
\ba{l}
z^{{\scriptscriptstyle (0)}}_{h}\!(x^0\!,Z)=x^0\!-\!\Xi_h^{-1}\!
\left( x^0\!-\! Z \right),\\[6pt]
 Z^{{\scriptscriptstyle (0)}}_{h}\!(x^0\!,z)=x^0\!-\!\Xi_h\!
\left( x^0\!-\! z \right)=z\!-\! Y^3_h(x^0\!-\! z),\\[6pt]
\bx^{{\scriptscriptstyle \perp(0)}}_{h}(x^0,\bX )=\bX^{{\scriptscriptstyle \perp}}
\!+\!\bY^{{\scriptscriptstyle \perp}}_h\!\left[x^0\!-\!
z^{{\scriptscriptstyle (0)}}_{h}\!(x^0\!,Z )\right],\\[6pt]
\bX^{{\scriptscriptstyle \perp(0)}}_{h}(x^0\!,\bx)=
\bx^{{\scriptscriptstyle \perp}} \!-\!
\bY\!^{{\scriptscriptstyle \perp}}_h\!\left(x^0\!-\!z\right).
\ea \label{hatxtxp}
\ee
The above functions fulfill in particular the following relations:
\be
\partial_0  Z ^{{\scriptscriptstyle (0)}}_h =-u_h^{{\scriptscriptstyle  z(0)}},
\qquad\quad\partial_z  Z ^{{\scriptscriptstyle (0)}}_h =\gamma_h^{{\scriptscriptstyle  (0)}}
,\qquad\quad\partial_Z z^{{\scriptscriptstyle (0)}}_h
=1/ {\widetilde{\gamma_h}^{{\scriptscriptstyle (0)}}},\qquad\quad
 \partial_Z \bx^{{\scriptscriptstyle \perp(0)}}_h=- 
\tbb_{h}^{{\scriptscriptstyle \perp(0)}}.                    \label{hatdtzz'}
\ee
\label{prop2}
\end{prop}

{\bf Proof} \ \ Solving the functional equations 
\be
x^0-z=\Xi_h^{-1}\left(x^0- Z\right) \qquad\qquad\Leftrightarrow \qquad\qquad x^0- Z =\Xi_h(x^0-z)
\label{txixi'}
\ee
resp. with respect to $z,Z$ we find the functions (\ref{hatxtxp})$_{1,2}$, which are the inverses of each 
other for each fixed $x^0$. Consider now the equation
\be
\bx-\bX=\bY\!_h(x^0\!-\!z).
\label{tXtX}
\ee
The $z$-component is equivalent to (\ref{txixi'})$_2$, by the equality
 $\Xi_h(\xi)\!=\!\xi\!+\!Y^3_h(\xi)$, so it is satisfied by (\ref{hatxtxp})$_{1,2}$.
Solving the transverse components resp. with respect to  $\bx,\bX$ we resp. find  (\ref{hatxtxp})$_{3,4}$.
which are clearly the inverse of each other for any fixed $x^0$. Moreover $x^0\!\le\!z$
implies $\bx\!=\!\bX$,  because $\bY_h(\xi)\!=\! 0$ for $\xi\!\le\!0$.
Deriving(\ref{hatxtxp})$_1$  with respect to  $x^0$ (holding $Z$ fixed) we  find with the help of
(\ref{n=0'+})$_6$,
\be
\partial_0z^{{\scriptscriptstyle (0)}}_{h}\!(x^0\!,Z)=1\!-\!
\frac 1{\gamma_h^{{\scriptscriptstyle (0)}}[\Xi_h^{-1}\!\!\left(x^0\!-\! Z\right)]}
=\beta_h^{{\scriptscriptstyle z(0)}}[\Xi_h^{-1}\!\!\left(x^0\!-\! Z\right)]
=\beta_h^{{\scriptscriptstyle z(0)}}[x^0\!-\!z^{{\scriptscriptstyle (0)}}_{h}\!(x^0\!,Z)]
=\tilde\beta_h^{{\scriptscriptstyle z(0)}}\!(x^0\!,Z)         \label{help}
\ee
i.e. the $z$-component of (\ref{lageul}) with the choice
$\bu_h\!=\!\bu_h^{{\scriptscriptstyle (0)}}$.
Deriving both sides of (\ref{tXtX}) with respect to  $x^0$ (holding $ \bX$ fixed) we find,
by the identity $\bY\!_h{}'\!=\!\bu^{{\scriptscriptstyle (0)}}_h$, eq. (\ref{help}) and (\ref{n=0'+})$_6$
$$
\partial_0 \bx^{{\scriptscriptstyle (0)}}_{h}=
\bY\!_h{}'[x^0\!-\!z^{{\scriptscriptstyle (0)}}_{h}\!(x^0\!,Z)]
\left[1\!-\!\partial_0 z^{{\scriptscriptstyle (0)}}_{h}\!( Z ,x^0 )\right]
=\frac{\bu^{{\scriptscriptstyle (0)}}_h }{\gamma^{{\scriptscriptstyle (0)}}_h}
[x^0\!-\!z^{{\scriptscriptstyle (0)}}_{h}\!(x^0\!,Z)]
=\tbb^{{\scriptscriptstyle (0)}}_h\!(Z ,x^0),
$$
i.e. $\bx^{{\scriptscriptstyle (0)}}_{h}( Z ,x^0)$
fulfills all components of eq. (\ref{lageul}) with the choice
$\bu_h\!=\!\bu_h^{{\scriptscriptstyle (0)}}$, as claimed.
Deriving (\ref{hatxtxp})$_2$ with respect to resp. $x^0,z$ (holding the other variable fixed) we immediately find (\ref{hatdtzz'})$_{1,2}$.
We find (\ref{hatdtzz'})$_3$ deriving (\ref{hatxtxp})$_1$ with respect to  $Z$ (holding $x^0$ fixed):
$$
\ba{l}
\partial_Zz^{{\scriptscriptstyle (0)}}_{h}\!(x^0\!,Z)=-\!
\frac 1{\gamma_h^{{\scriptscriptstyle (0)}}[\Xi_h^{-1}\!\!\left(x^0\!-\! Z\right)]}
=\frac 1{\gamma_h^{{\scriptscriptstyle (0)}}[x^0\!-\!
z^{{\scriptscriptstyle (0)}}_{h}\!(x^0\!,Z )]}.   
\ea
$$
We find  (\ref{hatdtzz'})$_4$  deriving (\ref{hatxtxp})$_3$    with respect to  $Z$ 
(holding $x^0$ fixed) and using (\ref{hatdtzz'})$_3$:
$$
\ba{l}
-\partial_Z \bx^{{\scriptscriptstyle \perp(0)}}_{h}(x^0\!,\!\bX )=\bu^{{\scriptscriptstyle \perp(0)}}_h\!\!\left[x^0\!-\!
z^{{\scriptscriptstyle (0)}}_{h}\!(x^0\!,\!Z )\right]\partial_Zz^{{\scriptscriptstyle (0)}}_{h}\!(x^0\!,\!Z)=
\frac {\bu^{{\scriptscriptstyle \perp(0)}}_h}{\gamma_h^{{\scriptscriptstyle (0)}}}\!\left[x^0\!-\!
z^{{\scriptscriptstyle (0)}}_{h}\!(x^0\!,\!Z )\right]=\tbb_{h}^{{\scriptscriptstyle \perp(0)}}\!(x^0\!,\!Z).   
\ea \qquad \sq
$$

\medskip
{\bf Remarks 2.} \ In Proposition \ref{prop2}
the only functions not given explicitly are the $\Xi_h^{-1}\!$; however, their graphs are obtained from those 
of the $\Xi_h$ just interchanging dependent and independent variable.
If $\Ba^{{\scriptscriptstyle \perp}}$ has an upper bound, then also the
 $\gamma_h^{{\scriptscriptstyle (0)}}$ defined by (\ref{n=0'+})$_6$  has one $\gamma_{h{\scriptscriptstyle M}}^{{\scriptscriptstyle (0)}}$;  \ 
by (\ref{hatdtzz'})$_2$ \  $J_h\!=\!\partial_z Z^{{\scriptscriptstyle (0)}}_h\!\in\![1,\gamma_{h{\scriptscriptstyle M}}^{{\scriptscriptstyle (0)}}]$  \ for all $x$, \ guaranteeing that the rquirement of  global invertibility 
of \ $\bX_h^{{\scriptscriptstyle (0)}}(x^0, \cdot):\bx\mapsto \bX$ \ is fulfilled.\footnote{That in a generic plasma
this is not always the case can be seen e.g. in \cite{Daw59}.}
Up to multiplication by $c$ rhs(\ref{txixi'})$_1$ is nothing but the proper time lapse $\tau_h(x^0,Z)$ 
 of the fluid element initially located in $z\!=\!Z$ from the time $Z/c$ when 
it is reached by the EM wave to time $x^0/c$. In fact changing integration 
variable \ $x^0{}'\mapsto y\!:=\!\Xi_h^{-1}\!\!\left(x^0{}'\!-\! Z\right)$, 
$$
c\,\tau_h(x^0,Z)\!=\!\int\limits^{x^0}_Z\!\!\frac{dx^0{}'}{\tilde \gamma_h^{{\scriptscriptstyle (0)}}\!(x^0{}'\!,\!Z)}
\!=\!\int\limits^{x^0}_Z\!\!\frac{dx^0{}'}{\gamma_h^{{\scriptscriptstyle (0)}}[x^0{}'\!-\!
z^{{\scriptscriptstyle (0)}}_{h}\!(x^0{}'\!,\!Z )]}
\!=\!\int\limits^{x^0}_Z\!\!\frac{dx^0{}'}{\gamma_h^{{\scriptscriptstyle (0)}}[\Xi_h^{-1}\!\!\left(x^0{}'\!-\! Z\right)]}
\!=\!\!\!\!\!\!\int\limits^{\Xi_h^{-1}\!(x^0\!-\! Z)}_0\!\!\!\!\!\!\! dy\!=\!\Xi_h^{-1}\!\!\left(x^0\!-\! Z\right).
$$
Another way to prove (\ref{hatxtxp})$_1$ is to note that, as $\tilde\gamma_h\!-\!\tilde u_h^z\!=\!\partial(x^0\!-\!z^{{\scriptscriptstyle (0)}}_{h})/\partial(c\tau_h)$, then 
$\tilde\gamma_h\partial_0(\tilde\gamma_h\!-\!\tilde u_h^z)\!=\!0$ amounts to \
$\partial^2(x^0\!-\!z^{{\scriptscriptstyle (0)}}_{h})/\partial(c\tau_h)^2\!=\!0$, \ what implies
that $x^0\!-\!z^{{\scriptscriptstyle (0)}}_{h}$ is a first degree polynomial in $c\tau_h$, more
precisely $x^0\!-\!z^{{\scriptscriptstyle (0)}}_{h}\!=\!c\tau_h$ by the assigned initial conditions.

\medskip
\noindent
From (\ref{hatxtxp}) it follows that  the longitudinal displacement
of the $h$-th type of particles with respect to their initial position $\bX$ at time $x^0$  is
\be
\Delta z_h^{{\scriptscriptstyle (0)}}\! ( x^0\!,\! Z ) :=
z_h^{{\scriptscriptstyle (0)}}\! ( x^0\!,\! Z )\!-\!Z\,=\, Y^3_h\!\left[\Xi_h^{-1}\!
\left( x^0\!-\! Z \right)\right].                 \label{displace}
\ee
By (\ref{solvn=0}) the evolution of $\bA^{{\scriptscriptstyle \perp}}$ amounts to a translation
of the graph of $\Ba^{{\scriptscriptstyle \perp}}$.
Its value $\check \Ba{{\scriptscriptstyle \perp}}\!:=\!\Ba{{\scriptscriptstyle \perp}}(\check \xi)$ 
at some point  $\check \xi$ 
reachs the particles initially located in $Z$ at the time $\check x_h^0(\check \xi,Z)$ such that 
\be
\check x_h^0\!-\!\check \xi=z_h^{{\scriptscriptstyle (0)}}\! (\check x^0\!,\! Z )
\!\stackrel{(\ref{hatxtxp})_1}{=}\!\check x_h^0\!-\!\Xi_h^{-1}\! \left[\check x_h^0\!-\! Z \right] 
\qquad\Leftrightarrow \qquad
\check x_h^0(\check \xi,Z)\!=\!\Xi_h(\check \xi)\!+\! Z,                        \label{chain}
\ee
in the position $z_h^{{\scriptscriptstyle (0)}} (\check x^0\!, Z )=
\Xi_h(\check \xi)\!+\! Z\!-\!\check \xi =Y^3_h(\check \xi)\!+\! Z $. The corresponding displacement 
of these particles is independent of $Z$ and equal to
\be
\zeta_h=
\Delta z_h^{{\scriptscriptstyle (0)}}\! \left[\check x_h^0 (\check \xi,Z), Z \right]=Y^3_h(\check \xi)
\label{checkzeta}.
\ee

Replacing (\ref{hatdtzz'}) in  (\ref{n_h}) one obtains as lowest order non-zero densities
\be
n_h^{{\scriptscriptstyle(0)}}(x^0\!, z)=
\widetilde{n_{h0}}\!\left[z\!-\!Y^3_h(x^0\!-\! z)\right]\,\gamma_h^{{\scriptscriptstyle(0)}}(x^0\!-\! z).
\label{lowestorderdensityh}
\ee



\bigskip
As said, (\ref{hatxtxp}) gives also the  motion 
of a {\it single} test particle of charge $q_h$ and mass $m_h$ with 
position $\bX$ and velocity $\0$ at sufficiently early time, i.e. before the EM wave arrives; 
one just ignores that the test particle
 can be thought of as a constituent of a (zero density) fluid.
By a suitable Poincar\'e transformation one can obtain
the solution $\bx_h(x^0)$ of the Cauchy problem consisting of
the relativistic equation of motion of a charged test particle 
under the action of  an arbitrary free transverse plane EM travelling-wave 
$\bE\!^{{\scriptscriptstyle\perp}}(x)=\Be\!^{{\scriptscriptstyle\perp}}( x^0\!-\! \bx\cdot\bee)$
$\bB\!^{{\scriptscriptstyle\perp}}=\bee\!\wedge\!\bE\!^{{\scriptscriptstyle\perp}}$
($\bee$ is the unit vector of the direction of propagation of the wave, and 
we no longer require $\bE\!^{{\scriptscriptstyle\perp}}, \bB\!^{{\scriptscriptstyle\perp}}$ to vanish for  $x^0\!-\! \bx\cdot\bee\!<\!0$) with {\it arbitrary initial conditions} $\bx_h(0)\!=\!\bx_0$, 
$\frac{d\bx_h}{dx^0}(0)\!=\!\bb_0$.\footnote{One first
finds the boost $B$ from the initial reference frame $\F$ to a new one $\F'$ where $\bb_0'\!=\!\0$
(this maps the transverse plane electromagnetic wave into a new one), then a rotation
$R$ to a reference frame $\F''$ where the 
plane wave propagates in the positive $z$-direction,
finally the translation $T$ to the reference frame $\underline\F$ where also $\underline\bx_0\!=\!\0$.
Naming $\underline x^\mu$ the spacetime coordinates and $\underline A^\mu,\underline F^{\mu\nu},...$ 
the fields with respect to   $\underline \F$, it is 
 $\frac{d\underline\bx_h}{d\underline x^0}(0)\!=\!\0$,  $\underline\bx_h(0)\!=\!\0$,
and $\underline\bE\!^{{\scriptscriptstyle\perp}}(x)\!=\!\underline\Be\!^{{\scriptscriptstyle\perp}}(\underline x^0{}\!-\!\underline z)$, \ 
$\underline\bB\!^{{\scriptscriptstyle\perp}}\!=\!\hat{{\underline{\bm z}}}\!\wedge\!\underline\bE\!^{{\scriptscriptstyle\perp}}$.
Since the part of the EM which is already at the right of the particle at  $\underline x^0\!=\!0$ will not 
come in contact with the particle nor affect its motion,
the solution $\underline x_h(\underline x^0)$ of the Cauchy problem with respect to 
$\underline\F$ does not change if we replace 
$\underline\Be^{{\scriptscriptstyle\perp}}\!(\underline x^0\!-\!\underline z)$ by the `cut' counterpart
$\Be_{{\scriptscriptstyle\theta}}^{{\scriptscriptstyle\perp}}\!(\underline x^0\!-\!\underline z)\!:=\!
\underline\Be^{{\scriptscriptstyle\perp}}\!(\underline x^0\!-\!\underline z)\,\theta(\underline x^0\!-\!\underline z)$
($\theta$ stands for the Heaviside step function).
Clearly 
$\Be^{{\scriptscriptstyle\perp}}_{{\scriptscriptstyle\theta}}(\xi)$ and 
$\Ba\!^{{\scriptscriptstyle\perp}}_{{\,\scriptscriptstyle\theta}}(\xi)\!:=\!\!\int^\xi_0\!\! d\xi' \,\Be_{{\scriptscriptstyle\theta}}^{{\scriptscriptstyle\perp}}\!(\xi')$
fulfill $\Be_{{\scriptscriptstyle\theta}}^{{\scriptscriptstyle\perp}}(\xi)\!=\!\Ba\!_{{\,\scriptscriptstyle\theta}}^{{\scriptscriptstyle\perp}}(\xi)\!=\!\0$
if $\xi\!\le\!0$; therefore, denoting as $\bu^{{\scriptscriptstyle (0)}}_{h{\scriptscriptstyle\theta}}(\xi),\bx^{{\scriptscriptstyle (0)}}_{h{\scriptscriptstyle\theta}}(\underline x^0\!,\bX ),...$ the functions of the previous section obtained choosing 
 $\Ba\!^{{\scriptscriptstyle\perp}}(\xi)\equiv\Ba\!_{{\,\scriptscriptstyle\theta}}^{{\scriptscriptstyle\perp}}(\xi)$, 
we find
$\underline \bx_h(\underline x^0)\!=\!\bx^{{\scriptscriptstyle (0)}}_{h{\scriptscriptstyle\theta}}(\underline x^0\!,\0 )$.
The solution in $\F$ is finally obtained applying the inverse Poincar\'e transformation $P^{-1}$ 
to $\underline \bx_h(\underline x^0)$.
}
%

%
\medskip
{\bf Bibliographical note.} \ Notably, (\ref{n=0'+}),  (\ref{hatxtxp}),
together with their just mentioned generalization to initial conditions $\bb_0\!\neq\!\0$,
agree with the parametric equations of motion of a test particle
obtained by resolution of its Hamilton-Jacobi equation  \cite{LanLif62,EbeSle68,EbeSle69}\footnote{
We thank the referees for pointing out such references. Formulae
(2) of p. 128 in \cite{LanLif62} are parametric equations (with parameter $\xi$) of the motion
after a choice of the space-time origin such that $\bx_h(0)\!=\!\0$,
but with arbitrary $\bb_0$; in the case $\bb_0\!=\!\0$ 
(corresponding to  ${\bm f}\!=\!\0$, $\gamma\!=\!1$ in the notation of \cite{LanLif62}) these
equations amount (in our notation) to the system of parametric equations 
$\bx\!=\!\bY(\xi)$, $\xi\!=\!x^0\!-\!z$, equivalent to (\ref{tXtX}) with $\bX\!=\!\0$.
In the literature we have found no analog of (\ref{hatxtxp}), which solve  (\ref{tXtX})
in terms of the function $\Xi_h$ and its inverse, and of (\ref{hatdtzz'}); in fact, both find their
natural context in the framework of fluid (plasma) physics.
}.  

\section{Integral equations for plane waves}
\label{integral}

In this section we reformulate the PDE's (\ref{lageul}), (\ref{Maxwell0}-\ref{hom'3})
with (\ref{asyc}) as integral equations. 

Let $\widetilde{N}_h(Z)$ be a primitive function of
$\widetilde{n_{h0}}(Z)$: \
$\widetilde{N}_h( Z ):=\int^{ Z } d Z' \widetilde{ n_{h0}}(Z')$;  \
$\widetilde{N}_h$ is defined up to an additive constant. 
 Setting \
$N_h(x^0,z)\!:=\!\widetilde{N}_h[ Z_h(x^0,z)]$, \ by (\ref{n_h}-\ref{j_h}) one easily shows\footnote{For instance,
$\partial_z N_h(x^0,z)\!=\!(\partial_Z\widetilde{N}_h)[Z_h(x^0 \!,z)]\,\partial_zZ_h
\!=\!\widetilde{n_{h0}}[Z_h(x^0 \!,z)]\,\partial_zZ_h 
\!=\!n_{h0}(x^0 \!,z)\,\partial_zZ_h \!=\!n_h(x^0 \!,z)$.}
\be
\partial_z N_h(x^0,z)\!=\!n_h(x^0 \!,z),\qquad\ \qquad\partial_0 N_h(x^0,z)\!=\!-(n_h\beta_h^z)(x^0 \!,z).\label{dN}
\ee
By (\ref{dN})$_1$ $N_h(x^0,z)$ is a primitive of \ $n_h(x^0\!,\!z)$ \
at fixed $x^0$.  There follows

\begin{prop} For any $\bar Z\!\in\!\mathbb{R}$ eq.
(\ref{Maxwell0}-\ref{Maxwell3}), (\ref{asyc}) are solved by
\be
  E^{{\scriptscriptstyle z}}(x^0,z)=4\pi \sum\limits_{h=1}^kq_h
\widetilde{N}_h[ Z_h(x^0\!,z)], \qquad
\qquad\widetilde{N}_h(Z):=\int^{Z}_{\bar Z}\!\!\! d Z'\,\widetilde{n_{h0}}(Z');
\label{expl}
\ee
the neutrality condition (\ref{asyc})$_3$ implies $\sum\limits_{h=1}^kq_h \widetilde{N}_h(Z)\equiv 0$.
\end{prop}
Formula  (\ref{expl}) gives the solution of
(\ref{Maxwell0}-\ref{Maxwell3})
explicitly in terms of the initial densities,
up to determination of the functions $ Z_h(x^0\!,z)$.

\begin{figure}[ht]
\begin{center}
\includegraphics[width=14cm]{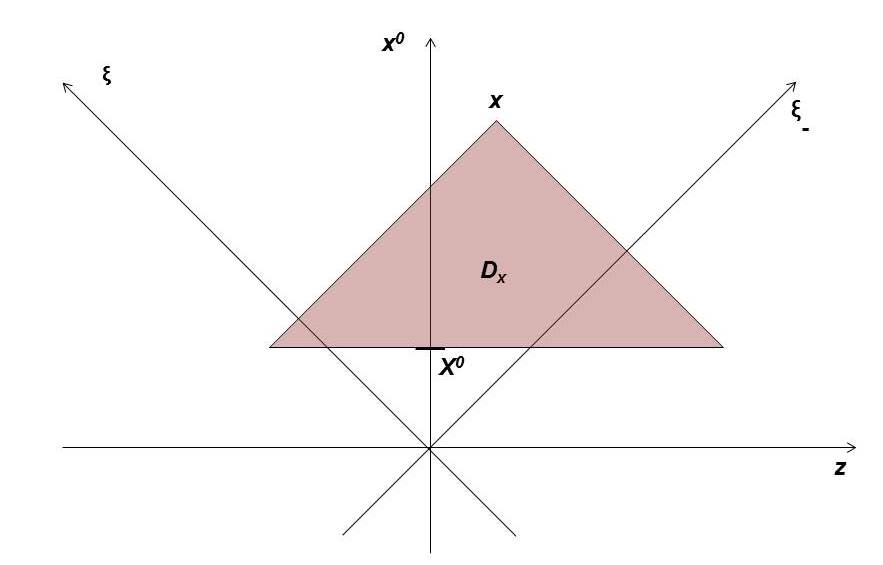}
\end{center}
\caption{}
\label{Dx}       
\end{figure}

\medskip
We recall that the Green function of the d'Alembertian
$\partial_0^2\!-\!\partial_z^2=4\partial_+\partial_-$ is $1/2$
the characteristic function of the causal cone
$T\!:=\!\{ x\:\:|\:\:  x^0  \!>\! |z| \}$, i.e. 
\be
G(x^0,z)=\frac 12 \theta(\xi) \theta(\xim)= \frac 12 \theta(x^0\!-\!z) \theta(x^0\!+\!z)=
 \frac 12 \theta(x^0\!-\!|z|),
\ee
where $\xim\!\!:=\!x^0\!\!+\!z$, \ 
$\partial_-\!:=\!\partial/\partial_{\xim}\!=\!\frac 12(\partial_0\!+\!\partial_z) $, \
$\theta$ is the Heaviside step function, so that
the general solution $y$ of an equation of the form \
$(\partial_0^2\!-\!\partial_z^2)y(x^0,z)=w(x^0,z)$ \  for $x^0\ge  X^0 $ \ is
\bea
y(x^0,z) &=& y_f(x^0,z)
+\displaystyle\int\limits^{\infty}_{  X^0 }\!\! dx^0{}'\!\!\!
\int\limits^{\infty}_{-\infty}\!\!\!dz'\, G\!\left( {x^0}\!-\!{x^0}',z\!-\!z'\right)\!
 w\left(x^0{}'\!,\!z'\right)\nn
&=& y_f(x^0,z)+ \frac 12   \displaystyle\int_{D\!_{x}}\!\!\!\!\!\! d^2x'
\, w\!\left({x^0}',z'\right)\qquad \qquad x^0\ge  X^0 ,
\label{Green1}
\eea
where $y_f(x^0,z)=y_+(x^0\!-\!z)+y_-(x^0\!+\!z)$,
 with arbitrary functions \ $y_+(\xi),y_-(\xim)$, and
\be
\ba{l}
D\!_{X^0\!,x}\!:=\!\{ x'\:|\:  X^0 \!\le\! {x^0}' \!\le\! x^0,
\, |z\!-\!z'|\!\le\!{x^0}\!-\!{x^0}' \}=\{ x'\:|\: 2 X^0 \!\le\!\xi'\!+\!\xim',
\, \xi'\!\le\! \xi, \, \xim'\!\le\! \xim\},   \label{defs}
\ea
\ee
is the isosceles triangle shown in fig. \ref{Dx}; if
$w(x^0,z)$ vanishes at early times (or as $x^0\to-\infty$) then (\ref{Green1})
holds also with $ X^0 =-\infty$.
The freedom in the choice of the ``pump'' $y_f$
amounts to the freedom in the assignment
of the initial (or asymptotic) conditions.

In the sequel we abbreviate
\be
\ba{l}
\varepsilon^z_h:=\frac{-q_h  E^z}{m_hc^2},\qquad\quad
Q:=\frac{2\pi }{c^2}\sum\limits_{h=1}^k
\frac{q_h^2n_h}{m_h \gamma_h}, \qquad\quad
\Rightarrow \qquad\quad
Q\bA\!^{{\scriptscriptstyle\perp}}=-2\pi \sum\limits_{h=1}^k
q_hn_h\bb^{{\scriptscriptstyle\perp}}_h;  \label{defs2}
\ea
\ee
$Q$ is the sum of the squared plasma frequencies of all species of particles, divided by $2c^2$.

\medskip
Henceforth we assume  that \underline{$n_{h}(0,z)\!=\!0$ for $z\!<\!0$}.
Then by (\ref{asyc}) for $x^0\!\le\! 0$ the EM wave is free and 
$\bA\!^{{\scriptscriptstyle\perp}}$ is of the form
$\bA\!^{{\scriptscriptstyle\perp}}( x^0 ,z)\equiv \Ba\!^{{\scriptscriptstyle\perp}}( x^0 \!-\!z)$,
with $\Ba\!^{{\scriptscriptstyle\perp}}(\xi)\!=\!0$  for  $\xi\!\le\! 0$, 
and we may choose 
$X^0\!=\!0$ in   (\ref{Green1})  [hence
$\widetilde{n_{h0}}(Z)\!\equiv\! n_{h}(0,\!Z)$]. Let $D\!_{x}\!=\!D\!_{0\!,x}$;
by (\ref{Green1}), eq. (\ref{Maxwell12}) equipped with such an initial condition
is equivalent to the integral equation
in the unknown $\bA\!^{{\scriptscriptstyle\perp}}$
\be
\bA\!^{{\scriptscriptstyle\perp}}(x^0\!,z)-
\Ba\!^{{\scriptscriptstyle\perp}}(x^0\!-\!z)=-\!\!
\displaystyle\int_{D\!_{x}}\!\!\!\!\!\! d^2x'\left[ Q\bA\!^{{\scriptscriptstyle\perp}}
\right]\!\!(x')= \!\!
\displaystyle\int\limits^{x^0}_{0}\!\! dx^0{}'\!\!\!\!\!\!\!
\int\limits^{{x^0}\!-\!{x^0}'\!+\!z}_{{x^0}'\!-\!{x^0}\!+\!z}\!\!\!\!\!dz'
\left[2\pi \sum\limits_{h=1}^k
q_hn_h\bb^{{\scriptscriptstyle\perp}}_h
\right]\!\!({x^0}', z');           \label{inteq1}
\ee

Let \ $s_h\!:=\!\gamma_h\!-\! u^z_h$; \ this is positive-definite. \
In the Lagrangian description
(\ref{bla'}) reads $\tilde\gamma_h\partial_0 \tilde s_h=\tilde s_h\tilde\varepsilon^z_h\!+\!
 \widetilde{\partial_-\bu_h^{{\scriptscriptstyle\perp}}{}^2}$;
the Cauchy problem with initial condition $\tilde s_{h0}\!\equiv\! 1$ is
equivalent\footnote{Given $\tilde\varepsilon^z_h$, \ $\tilde \sigma_h(x^0, Z ):=
e^{\int\limits^{x^0}_{ 0 }\!\!dx^0{}'\tilde\varepsilon^z_h(x^0{}', Z )}$
fulfills  the homogenous equation
$\partial_0 \tilde \sigma_h=\tilde \sigma_h\tilde\varepsilon^z_h$. Looking for
$\tilde s_h$ in the form $\tilde s_h=\tilde \sigma_h\tilde f_h$ one finds that it must be
 $\partial_0 \tilde f_h=
\frac {\widetilde{\partial_-\bu_h^{{\scriptscriptstyle\perp}}{}^2}}{\tilde \sigma_h\tilde\gamma_h}$;
integrating over $[ 0 ,x^0]$ and imposing the initial condition one finds
$f_h$, which replaced in the Ansatz \ $\tilde s_h=\tilde \sigma_h\tilde f_h$ \
gives (\ref{inteq2}).} to the integral equation
\be
\ba{lll}
\tilde s_h &:=& \tilde \gamma_h\!-\! \tilde u^z_h= e^{\int\limits^{x^0}_{ 0 }\!\!dx^0{}'\frac {\tilde\varepsilon^z_h}{\tilde\gamma_h}(x^0{}', Z )}
\!\!\!+\!\!\displaystyle\int\limits^{x^0}_{ 0 }\!\!
dx^0{}'\,e^{\int\limits^{x^0}_{x^0{}'}\!\!dx^0{}''\frac {\tilde\varepsilon^z_h}{\tilde\gamma_h}(x^0{}'', Z )} \left[\frac { \widetilde{\partial_-\bu_h^{{\scriptscriptstyle\perp}}{}^2}}{\tilde\gamma_h}\right]\! (x^0{}', Z ). 
 \ea   \label{inteq2}
\ee
It is straightforward to show that $u_h^z,\gamma_h,\bb_h^{{\scriptscriptstyle\perp}},\beta_h^z$ are recovered from $s_h,\bu_h^{{\scriptscriptstyle\perp}}$ through the formulae
\be
\ba{ll}
\displaystyle\gamma_h\!=\!\frac {1\!+\!\bu_h^{{\scriptscriptstyle\perp}}{}^2\!\!+\!s_h^2}{2s_h}, \qquad\qquad & \displaystyle\bb_h^{{\scriptscriptstyle\perp}}\!=\! \frac{\bu_h^{{\scriptscriptstyle\perp}}}{\gamma_h} \!=\!\frac{2s_h\bu_h^{{\scriptscriptstyle\perp}}}
 {1\!+\!\bu_h^{{\scriptscriptstyle\perp}2}\!\!+\!s_h^2 },\\[18pt]
\displaystyle u_h^z\!=\!\frac {1\!+\!\bu_h^{{\scriptscriptstyle\perp}}{}^2\!\!-\!s_h^2}{2s_h}, 
 \qquad\qquad & \displaystyle
\beta_h^z\!=\! \frac{u_h^z}{\gamma_h}\!=\!\frac{1\!+\!\bu_h^{{\scriptscriptstyle\perp}2}\!
\!-\!s_h^2 } {1\!+\!\bu_h^{{\scriptscriptstyle\perp}2}\!\!+\!s_h^2 }.  \label{u_hs_h}
\ea
\ee
The Cauchy problem (\ref{lageul}) with initial condition \ $\bx_{h}(0,\bX)\!=\!\bX$ \
 is equivalent to the integral equations
\be
z_h(x^0\!,Z)= Z \!+\!\! \displaystyle\int\limits^{x^0}_{0}\!\!
dx^0{}'\, \beta_h^z[x^0{}'\!,\! z_h(x^0{}'\!,\!Z)],\qquad
\bx^{{\scriptscriptstyle \perp}}_{h}(x^0\!,\!\bX )=\bX^{{\scriptscriptstyle \perp}}
\!+\!\! \displaystyle\int\limits^{x^0}_{
0}\!\!
dx^0{}'\, \bb_h^{{\scriptscriptstyle \perp}}[x^0{}'\!,\! z_h(x^0{}'\!,\!Z)]\qquad \label{inteq0}
\ee

Summing up, we have shown that making use of  (\ref{hom'12}), (\ref{n_h}), (\ref{expl}),
(\ref{u_hs_h}) the determination of the evolution of the system is reduced
to solving the system of integral equations  (\ref{inteq1}),  (\ref{inteq2}) and (\ref{inteq0})$_1$
 in the unknowns $\bA\!^{{\scriptscriptstyle\perp}}, s_h, z_h$
[note that, once this is solved, (\ref{inteq0})$_2$ becomes known].

\section{Impact of a short pulse on a step-density plasma}
\label{constdensity}

\begin{figure*}[ht]
\begin{center}
\includegraphics[width=8cm]{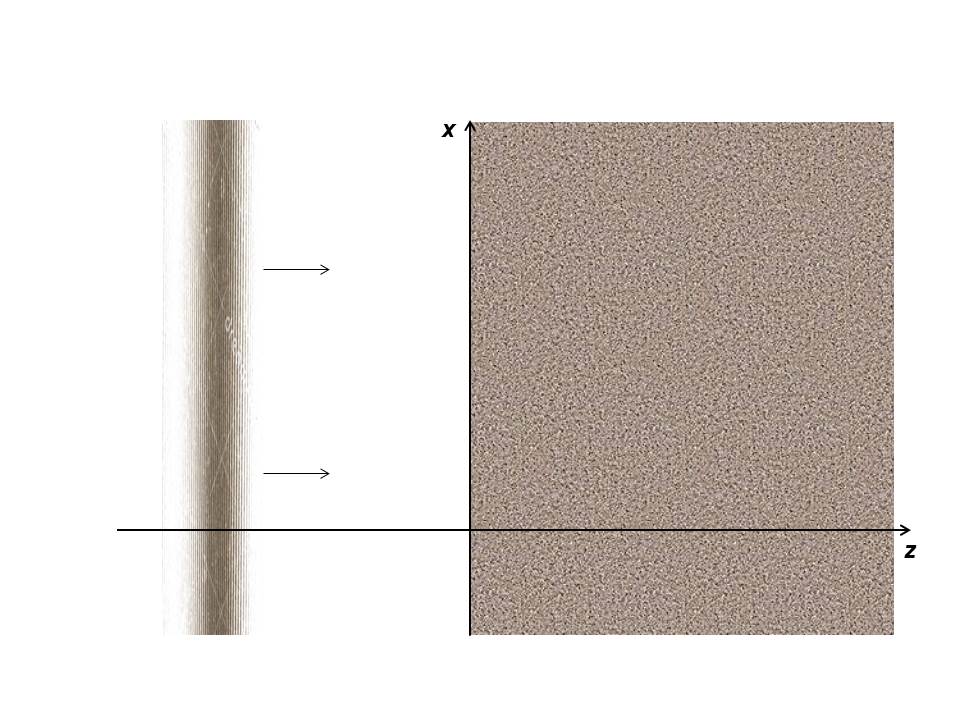}\hfill
\includegraphics[width=8cm]{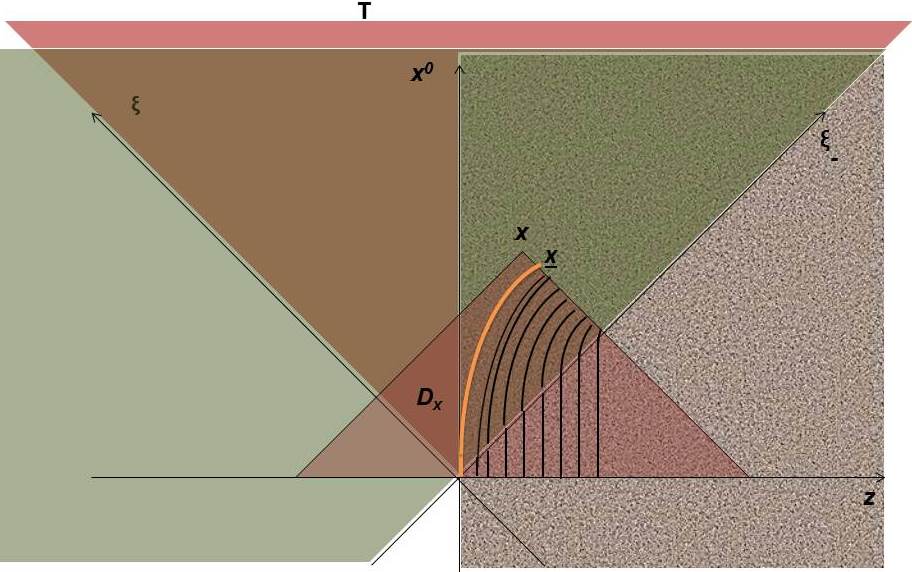}
\end{center}
\caption{}
\label{trajectories4}       
\end{figure*}

In the present section  we assume in addition that $\widetilde{n_{h0}}( Z )$ 
are zero for $Z\!<\!0$ and constant for $Z\!>\!0$:
 $\widetilde{n_{e0}}( Z )\!=\!n_0\theta(Z)$, etc, as depicted in fig. \ref{trajectories4}-left. 
Moreover, we are interested in studying the equations for
so small $x^0$ (small times after the beginning of the interaction)
that the motion of ions can be neglected. Hence we consider ions as infinitely massive, so that they remain at rest [$Z_h(x^0\!,\! z)\!\equiv\! z$ for $h\!\neq\! e$], 
have constant densities, and their contribution  to
$Q\bA\!^{{\scriptscriptstyle\perp}}$ disappears; only  electrons  contribute:
$Q\bA\!^{{\scriptscriptstyle\perp}}\!=\! -2\pi e n_e\bb^{{\scriptscriptstyle\perp}}_e$. 
Choosing $\bar Z\!=\!0$ \  in (\ref{expl}) we find
\bea
 E^{{\scriptscriptstyle z}}(x^0\!,z)=4\pi \sum\limits_{h=1}^kq_h
\widetilde{N}_h[ Z_h(x^0\!,z)]= 4\pi e n_0\left\{
z\,\theta(z)\!-\! Z_e(x^0\!, z)\,\theta[ Z_e(x^0\!, z)]\right\}.         \label{elField}
\eea
If $z,Z\!>\!0$ this implies the well-known result (see e.g. \cite{Daw59,AkhPol67}) 
that  at $x^0$ the electric field acting  on the electrons initally located
in $Z$  is \ $\widetilde{E}^{{\scriptscriptstyle z}}(x^0\!,Z)\!=\! 4\pi e n_0
\Delta z_e(x^0\!, Z)$, \ i.e. proportional to the displacement \ $\Delta z_e(x^0\!, Z)\!:=\! z_e(x^0\!, Z)\!-\! Z$ \
with respect to their initial position. The system of integral equations  (\ref{inteq1}),
 (\ref{inteq2}), (\ref{inteq0})$_1$ to be solved takes the form 
\bea
\bu_e^{{\scriptscriptstyle\perp}}(x^0\!,z)-\bu_e^{{\scriptscriptstyle\perp(0)}}(x^0\!-\!z)=
-\frac{2\pi e^2}{m_ec^2} \!\! \displaystyle\int_{D\!_{x}}\!\!\!\! d^2x'\:
\left[n_e\bb^{{\scriptscriptstyle\perp}}_e \right]\!\!(x')
\nn  \tilde s_e = e^{\tilde r_e}
\!+\!\!\displaystyle\int\limits^{x^0}_{ 0 }\!\!
dx^0{}' \left[e^{-\tilde r_e} \frac { \widetilde{\partial_-\bu_e^{{\scriptscriptstyle\perp}2}}}
{\tilde\gamma_e}\right] (x^0{}'\!, Z ), \label{intsystem}\\
\Delta z_e(x^0,Z):=z_e(x^0,Z)- Z =\!\! \displaystyle\int\limits^{x^0}_{0}\!\!
dx^0{}'\,\tilde \beta_e^{{\scriptscriptstyle z}}({x^0}', Z )\nonumber
\eea
 ($Z\ge 0$), where $\bb_e $ is related to the unknowns 
$\bu^{{\scriptscriptstyle\perp}}_e,s_e,z_e$ by (\ref{u_hs_h}) and, because of  (\ref{n_h}),  (\ref{elField})
\be
n_e\!=\!n_0\,\theta[Z_e]\,\partial_zZ_e,\qquad
\tilde r_e(x^0\!,\! Z )\!:=\!\int\limits^{x^0}_{ 0 }\!\!dx^0{}'\frac {\tilde\varepsilon^z_e}{\tilde\gamma_e}
(x^0{}'\!\!,\! Z )\!=\!4K\!\!\int\limits^{x^0}_{Z}\!\!dx^0{}'\frac {z_e\theta[z_e]\!-\! Z\theta(Z)}{\tilde\gamma_e}
(x^0{}'\!\!,\! Z ),                                \label{defr}
\ee
where \ $K\!:=\!\frac{\pi e^2 n_0}{m_ec^2}$ \ 
($K$ is the square of the equilibrium plasma frequency, divided by $4c^2$). 
As \  $n_e\bb^{{\scriptscriptstyle\perp}}_e\!\propto\! 
n_e\bA^{{\scriptscriptstyle\perp}}\!=\!0$
\ outside $T$, one can replace $D\!_{x}$ by $D\!_{x}\cap T$ in  (\ref{intsystem})$_1$.

\noindent
For $n_0\!=\!0$ the solution of (\ref{intsystem}) is $(\bu^{{\scriptscriptstyle\perp}}_e,
\tilde s_e,z_e)=
(\bu^{{\scriptscriptstyle\perp(0)}}_e,1,z_e^{{\scriptscriptstyle(0)}})$. 
For $n_0\!>\!0$ we can approximate better and better the solution by an iterative 
procedure:  replacing 
the approximation after $k$ steps [which we will distinguish by the superscript $(k)$]
at the rhs of (\ref{intsystem}) we will
obtain at the left-hand side (lhs) the approximation after $k\!+\!1$ steps \cite{Fioetal}. 
Here we  stick to the first step ($k\!=\!0$), 
\bea
\bu_e^{{\scriptscriptstyle\perp(1)}}(x^0\!,z)-\bu_e^{{\scriptscriptstyle\perp(0)}}(x^0\!-\!z)=-\frac{2\pi e^2}{m_ec^2} \!\!
\displaystyle\int_{D\!_{x}}\!\!\!\! d^2x'\:\left[n_e^{{\scriptscriptstyle(0)}}\bb^{{\scriptscriptstyle\perp(0)}}_e \right]\!\!(x')
\nn \tilde s_e^{{\scriptscriptstyle(1)}} = e^{\tilde r_e^{{\scriptscriptstyle(0)}}}
\!+\!\!\displaystyle\int\limits^{x^0}_{ 0 }\!\!
dx^0{}'\,\left[e^{-\tilde r_e^{{\scriptscriptstyle(0)}}} 
\frac { \widetilde{\partial_-\bu_e^{{\scriptscriptstyle\perp(0)}2} }}
{\tilde\gamma_e^{{\scriptscriptstyle(0)}}}\right]\!  (x^0{}'\!, Z )=e^{\tilde r_e^{{\scriptscriptstyle(0)}}}, \label{appr1}\\
\Delta z_e^{{\scriptscriptstyle(1)}}(x^0,Z):=z_e^{{\scriptscriptstyle(1)}}(x^0,Z)-Z
= \!\! \displaystyle\int\limits^{x^0}_{0}\!\!
dx^0{}'\,\tilde \beta_e^{{\scriptscriptstyle z(1)}}({x^0}', Z ),\nonumber
\eea
and investigate for how long 
$(\bu^{{\scriptscriptstyle\perp(1)}}_e,\tilde s^{{\scriptscriptstyle\perp(1)}}_e,z_e^{{\scriptscriptstyle(1)}})$ remains ``close'' to $(\bu^{{\scriptscriptstyle\perp(0)}}_e,1,z_e^{{\scriptscriptstyle(0)}})$. In (\ref{appr1}) the tilde is to be understood as the change from the Eulerian to the  Lagrangian description 
performed approximating $\bx_e$ by $\bx_e^{{\scriptscriptstyle(0)}}$, and in the second line
we have used $ \partial\!_{{\scriptscriptstyle -}}\bu_e^{{\scriptscriptstyle\perp(0)}2}\!=\!0$. 
The electron density obtained approximating
$z_e$ by $z_e^{{\scriptscriptstyle(0)}}$, i.e. replacing (\ref{hatdtzz'}) in  (\ref{defr})$_1$
or equivalently $\widetilde{n_{e0}}( Z )\!=\!n_0\theta(Z)$ in (\ref{lowestorderdensityh}),
is
\be
n_e^{{\scriptscriptstyle(0)}}(x^0\!, z)=n_0\,\theta\left[z\!-\!Y^3_e(x^0\!-\! z)\right]\,\gamma_e^{{\scriptscriptstyle(0)}}(x^0\!-\! z).
\label{lowestorderdensity}
\ee

In the appendix we prove that for $Z\ge 0$
 we can give $\tilde r_e^{{\scriptscriptstyle(0)}}$ the more explicit form
\be
\tilde r_e^{{\scriptscriptstyle(0)}}(x^0\!, Z )\, =\, 
4K\,V_e^3\!\left[\Xi_e^{-1}\!( x^0\!\!-\! Z)\right],\qquad \qquad\mbox{where }\quad V_e^3(\xi):=
\displaystyle\int ^{\xi}_0\!\! dy\, Y^3_e(y).                \label{tilder}
\ee
As $Y^3_e(\xi)$ is strictly increasing,  $V_e^3(\xi),r_e^{{\scriptscriptstyle (0)}}(\xi)$ 
are not only strictly increasing, 
but also convex in all $]0,l[$. This implies $V_e^3(\xi)/Y_e^3(\xi)\!<\!\xi/2$ at least.
%

Now assume that $\bE^{{\scriptscriptstyle \perp}}(x^0\!,z)\!=\!\Be^{{\scriptscriptstyle \perp}}(x^0\!-\!z)$
with $\Be^{{\scriptscriptstyle \perp}}(\xi)\!=\!\epsilon_s(\xi)\Be_o^{{\scriptscriptstyle \perp}}(\xi)$,
where the envelope amplitude $\epsilon_s(\xi)\!\ge\! 0$  has a finite support $[0,l]$ \ 
(as depicted in fig. \ref{trajectories4}-left) and
slowly varies on the scale $\lambda \!\ll\! l$, and
$\Be_o^{{\scriptscriptstyle \perp}}(\xi)$ is a sinusoidally oscillating transverse vector  with period
$\lambda\!:=\!2\pi/k$. For the sake of definiteness, we shall consider
\be
\ba{lll}
\Be_o^{{\scriptscriptstyle \perp}}(\xi)\!=\! {\hat\bx}\cos k\xi,      
\quad & \Be_p^{{\scriptscriptstyle \perp}}(\xi)\!=\! 
{\hat\bx}\sin k\xi          \quad &\mbox{(linear polarization), or}\\[8pt]
\Be_o^{{\scriptscriptstyle \perp}}(\xi)\!=\!{\hat\bx}\cos k\xi\!+\!{\hat\by}\sin k\xi, 
\quad & \Be_p^{{\scriptscriptstyle \perp}}(\xi)\!=\! 
{\hat\bx}\sin k\xi\!-\!{\hat\by}\cos k\xi \quad &\mbox{(circular polarization).}
\ea
\ee
Then \ $\Ba^{{\scriptscriptstyle \perp}}(\xi)\!\simeq\! 
\frac{1}{k}\epsilon_s\Be_p^{{\scriptscriptstyle \perp}}(\xi)$, \ 
and, setting \ $w\!:=\!\frac{ e}{kmc^2}\epsilon_s$, 
 \ $\bu_e^{{\scriptscriptstyle \perp(0)}}(\xi)\!\simeq\! w(\xi)\Be_p^{{\scriptscriptstyle \perp}}(\xi)$, \ $\bY_e\!^{{\scriptscriptstyle \perp}}\!(\xi)
\!\simeq\! -\frac 1k w(\xi)\Be_o^{{\scriptscriptstyle \perp}}(\xi)$, \   $u_e^{{\scriptscriptstyle z(0)}}(\xi)\!\simeq\! \frac 12 w^2(\xi)\Be_p^{{\scriptscriptstyle \perp 2}}(\xi)$. \ 
Denoting the first (possibly unique) maximum point of \ $\Ba^{{\scriptscriptstyle \perp}2}$ \ as  \ $\xi_0$, \
in the appendix we give the rhs(\ref{appr1})$_1$ the more explicit expression (\ref{budiff}) and
 prove for \ $ 0\!\le\! \xi\!\le\! \xi_0$ and either polarization 
the inequality
\be
\frac{|\bA^{{\scriptscriptstyle \perp(1)}}(\xi,\xim)\!-\!
\Ba^{{\scriptscriptstyle \perp}}(\xi)|}{\epsilon_s(\xi)/k}=
\frac{|\bu_e^{{\scriptscriptstyle \perp(1)}}(\xi,\xim)\!-\!
\bu_e^{{\scriptscriptstyle \perp(0)}}(\xi)|}{w(\xi)}\le 
 \frac{K\lambda\xim}{2\pi }
. \label{budiffbound}
\ee
The lhs  is the difference 
between $\bu_e^{{\scriptscriptstyle \perp(1)}},\bu_e^{{\scriptscriptstyle \perp(0)}}$
(or, equivalently, the difference between $\bA^{{\scriptscriptstyle \perp(1)}},\Ba^{{\scriptscriptstyle \perp}}$)
normalized by their modulating envelope amplitude; if the polarization is circular then $|\Be_p^{{\scriptscriptstyle \perp}}|\!=\!1$, and the lhs is exactly the modulus
of the relative difference 
between $\bu_e^{{\scriptscriptstyle \perp(1)}},\bu_e^{{\scriptscriptstyle \perp(0)}}$
(or, equivalently, $\bA^{{\scriptscriptstyle \perp(1)}},\Ba^{{\scriptscriptstyle \perp}}$). 
If $\xim\!\ll\!  \frac{2\pi} {K\lambda}$ then lhs$\le$rhs$\ll\! 1$. 
So we can approximate 
\be
\bu_e^{{\scriptscriptstyle \perp}}\!\simeq\!\bu_e^{{\scriptscriptstyle \perp(1)}}\!\simeq\!\bu_e^{{\scriptscriptstyle \perp(0)}} \qquad \quad 
\mbox{in the space-time region }\:\: 0\!\le\!\xi\!\le\! \xi_0,\quad
0\!\le\!\xim\!\ll\!  \frac{2\pi} {K\lambda}.    \label{stregion}
\ee
Consequently and by (\ref{u_hs_h}), in the region (\ref{stregion}) we find by some computation
\bea
&& \beta_e^{{\scriptscriptstyle z(1)}}\!=\!\frac {1\!+\!\bu_e^{{\scriptscriptstyle\perp(1)}2}
\!\!-\!s^{{\scriptscriptstyle (1)}2}_e}{1\!+\!\bu_e^{{\scriptscriptstyle\perp(1)}2}\!\!+
\!s^{{\scriptscriptstyle (1)}2}_e }
\!\simeq\!\frac {1\!+\!\bu_e^{{\scriptscriptstyle\perp(0)}2}
\!\!-\!e^{2r^{{\scriptscriptstyle (0)}}_e}}{1\!+\!\bu_e^{{\scriptscriptstyle\perp(0)}2}\!\!+
\!e^{2r^{{\scriptscriptstyle (0)}}_e} }
=\frac {2(1\!+\!2u_e^{{\scriptscriptstyle z(0)}})}{1\!+\!2u_e^{{\scriptscriptstyle z(0)}}
\!\!+\!e^{2r^{{\scriptscriptstyle (0)}}_e}}\!-\!1,                      \qquad   \label{beta1}\\
&&\Delta z_e^{{\scriptscriptstyle (1)}}(x^0\!,\! Z )=\!\displaystyle\int\limits^{x^0}_{ Z }\!\!
dx^0{}'\,\tilde \beta_e^{{\scriptscriptstyle z(1)}}(x^0{}'\!\!,\! Z)\simeq
\!\!\!\!\!\!\!\displaystyle\int\limits^{\Xi_e^{-1}\!( x^0\!\!-\! Z)}_0
\!\!\!\!\!\!\!\! dy\, [\gamma_e^{{\scriptscriptstyle (0)}}\beta_e^{{\scriptscriptstyle z(1)}}](y)
,\qquad\\
&& 0\le [z_e^{{\scriptscriptstyle (0)}}\!-\!z_e^{{\scriptscriptstyle (1)}}](x^0\!,\! Z )
\simeq G[\Xi_e^{-1}\!( x^0\!\!-\! Z)], \qquad 
G(\xi)\!:=\!\!\int\limits^{\xi}_0\!\!\! dy\, g(y), \quad
g:=\frac{\left(\!1\!+\!2u_e^{{\scriptscriptstyle z(0)}}\!\right)\! 
\left(\!e^{2r^{{\scriptscriptstyle (0)}}_e}\!\!\!-\!1\!\!\right) }
{1\!+\!2u_e^{{\scriptscriptstyle z(0)}}
\!+\!e^{2r^{{\scriptscriptstyle (0)}}_e}},\qquad\\[8pt]
&& 0\le \frac{\Delta z_e^{{\scriptscriptstyle (0)}}\!-\!\Delta z_e^{{\scriptscriptstyle (1)}}}
{\Delta z_e^{{\scriptscriptstyle (0)}}}(x^0\!,\! Z )=
\frac{ z_e^{{\scriptscriptstyle (0)}}(x^0\!,\! Z )\!-\!z_e^{{\scriptscriptstyle (1)}}(x^0\!,\! Z )}
{ z_e^{{\scriptscriptstyle (0)}}(x^0\!,\! Z )\!-\! Z}
\simeq T[\Xi_e^{-1}\!( x^0\!\!-\! Z)],\qquad\qquad
T\!:=\!\frac{G}{Y_e^3}.\qquad\label{reldif}
\eea{}

By (\ref{reldif}), $T[\Xi_e^{-1}\!( x^0\!\!-\! Z)]$ gives the relative difference 
between the displacement $\Delta z_e$ in the zero-density and in the first corrected approximation. 
Hence the approximation \ $z_e(x^0\!,\! Z )\!\simeq\! z_e^{{\scriptscriptstyle(1)}}(x^0\!,\! Z )\!\simeq\! z_e^{{\scriptscriptstyle(0)}}(x^0\!,\! Z )$
may be good only as long as \ $T[\Xi_e^{-1}\!( x^0\!\!-\! Z)]\!\ll\! 1$. By (\ref{chain}),  the maximum $\Ba^{{\scriptscriptstyle \perp}}(\xi_0)$ reaches  the electrons initially located in $Z$ at the time $\bar x^0(Z)\!=\!\Xi_e(\xi_0)\!+\! Z$;  
therefore the approximation \ $z_e(x^0\!,\! Z )\!\simeq z_e^{{\scriptscriptstyle(0)}}(x^0\!,\! Z )$
may be good for all $x^0\!\le\!\bar x^0(Z)$  if
\be
T(\xi)\ll 1\qquad\quad 0\le\xi\le\xi_0,
\qquad\qquad\qquad
2Y^3_e(\xi_0)\!+\!\xi_0\!+\! 2Z \ll  \frac{2\pi} {K\lambda}.          \label{condgood}
\ee
The condition at the right guarantees that (\ref{stregion}) is fulfilled, 
since $\bar x^0\!+\!z_e^{{\scriptscriptstyle (0)}} (\bar x^0\!, Z )
\!=\!2Y^3_e(\xi_0)\!+\!\xi_0\!+\! 2Z$.
 In $[0,\xi_0]$ $g(\xi)$  is strictly increasing  
\footnote{$g(\xi)$ is strictly increasing as it can be put in the following form,
where the numerator is a strictly
increasing function and the denominator the sum of a strictly
decreasing function and a decreasing function in $[0,\xi_0]$:
$$
g(\xi)\!=\!\frac{1\!-\!e^{-2r^{{\scriptscriptstyle (0)}}_e(\xi)} }
{e^{-2r^{{\scriptscriptstyle (0)}}_e(\xi)}\!+\!
\frac 1{1+2u_e^{{\scriptscriptstyle z(0)}}\!(\xi) }}.
$$
},
hence $G(\xi)$ is strictly increasing and convex. Moreover,  as $\xi\!\to\!0$
$g(\xi)\!\simeq\! 2r^{{\scriptscriptstyle (0)}}_e(\xi)$, whence
$G(\xi)\!<\!g(\xi)\xi/2\!\lesssim\!4K\xi V^3_e(\xi)\!<2K\xi^2 Y^3_e(\xi)$ at least,
and we find not only $g(0)\!=\!G(0)\!=\!0$, but also $T(0)\!=\!0$,  as expected.
In fig. \ref{Plots}  we plot $T(\xi)$ for a gaussian pulse and several values of $K$; 
it is a strictly increasing, convex function.

If (\ref{condgood}) is fulfilled then 
by (\ref{checkzeta}) the displacement at time $\bar x^0$ of the electrons initially located
at $Z=Z$  is independent  of $Z$ and approximately equal to
\be
\zeta(Z)\simeq Y^3_e(\xi_0)=\frac 12 \xi_0u_{e\,m}^{{\scriptscriptstyle \perp}2}=\!\!
\int\limits^{\xi_0}_0\!\!\! d\xi\, u_e^{{\scriptscriptstyle z(0)}}(\xi)=\!\!
\int\limits^{\xi_0}_0\!\!\! d\xi\, \frac p2 
w^2\!(\xi),
\qquad p\!=\!\left\{\!\!\ba{ll}1\: & \mbox{circ. pol.}\\ 
\frac 12\: & \mbox{lin. pol.};\ea\right.
\label{zeta}
\ee
here $p\!\simeq\!1/2$ for linear polarization comes 
from the mean of \ $\sin^2\! k\xi$ \  over a period \  $\lambda$, \  and
$u_{e\,m}^{{\scriptscriptstyle \perp}2}$ stands for the mean of 
$\bu_{e}^{{\scriptscriptstyle \perp}2}$ over $[0,\xi_0]$ (if $w$ is symmetric around 
a unique maximum at $\xi_0=l/2$, the latter coincides with the mean over all  $[0,l]$). 

By (\ref{beta1}), the monotonicity of $r_e^{{\scriptscriptstyle (0)}}$ and the fact that \
$u_e^{{\scriptscriptstyle z(0)}}(\xi)\!\propto\! w^2(\xi)\!\to\! 0$ \ as \ $\xi\!\to\! l$,
there exists (at least) a point in $]\xi_0,l]$ where $\beta_e^{{\scriptscriptstyle z(1)}}$ vanishes;
let $\xi_1$ be the first one. Namely, in the first corrected approximation
the electrons at some point invert their motion, as we expect due to the restoring electric
back-force exerted by the ions.
The condition  $\beta_e^{{\scriptscriptstyle z(1)}}(\xi_1)\!=\!0$ amounts to
$1\!+\!2u_e^{{\scriptscriptstyle z(0)}}\!(\xi_1)=\exp[8KV_e^3(\xi_1)]$. Fixing $\xi_1$,
this can be solved for $K$ (which is physically tunable by the choice of $n_0$):
\be
K\equiv K(\xi_1)=\frac{\ln[1\!+\!2u_e^{{\scriptscriptstyle z(0)}}\!(\xi_1)]}{8V_e^3(\xi_1)}
=\frac{\ln[1\!+\!pw^2\!(\xi_1)]}{8V_e^3(\xi_1)};                     \label{Kxi1}
\ee
Manifestly, $K(\xi_1)$ is a strictly decreasing function from $K(0)\!=\! \infty$ to $K(l)\! =\! 0$. \ 
The displacement (\ref{zeta}) will be approximately the maximal one if 
 $0\!<\!(\xi_1\!-\!\xi_0)/\xi_0\!\ll\! 1$, while  (\ref{condgood}) is respected. 
To check these conditions one can proceed by
``trial and error'': choosing $\xi_1$ close to $\xi_0$, computing $K$ and checking whether 
(\ref{condgood}) is fulfilled. If so, (\ref{zeta}) is reliable for that value, and all smaller values,
of $K$. If not, 
one has to try with a larger $\xi_1$, which will give a smaller $K$ and a smaller $T(\xi)$ for $\xi\!\in\![0,\xi_0]$, until the maximum $T(\xi)$ [tipically, $T(\xi_0)$] is small enough.


\subsection{The slingshot effect}
\label{Sling}

The basic formula (\ref{zeta}) for the displacement is used in \cite{FioFedDeA13} to predict
and estimate the  {\it slingshot effect}, i.e. the expulsion in the negative $z$-direction 
of the plasma eletrons initially located around $Z\!=\!0$, shortly after the impact of
a suitable ultra-short and ultra-intense laser pulse in the form of a pancake 
normally onto a plasma.  The mechanism is very simple: the plasma electrons in 
a thin layer - just beyond the surface of the plasma - first are given sufficient 
electric potential energy by the displacement  (\ref{zeta})  with respect to the ions, 
then after the pulse are pulled back by the longitudinal 
electric force exerted by the latter and may leave the plasma. 
Sufficient conditions for this to happen are that the pancake is sufficiently thin  ($R\!\gg l$), its radius $R$
is not too small  ($R\!\gtrsim\!\zeta$), the EM field inside is sufficiently intense, and
the electron density $n_0$ is sufficiently low.
$R\!\gg l$ guarantees that the plane wave solutions considered in the previous sections
are sufficiently accurate in the plasma, 
especially in the forward boost phase, in the internal part of a cylinder of radius $R$.  
$R\!\gtrsim\!\zeta$ avoids that the way out of the mentioned thin layer of electrons 
within such a cylinder is  blocked by the electrons initially located just outside 
the cylindrical surface (which are attracted and move towards the cylinder axis).
The high intensity of the  EM field and the sufficiently low plasma density 
are needed for the longitudinal electric force to induce the back-acceleration of the electrons 
{\it after}  the pulse maximum has overcome them, in phase with the negative
ponderomotive force exerted by the pulse in its decreasing stage. Actually we impose
the stronger condition that $n_0$ is sufficiently low in order that 
(\ref{condgood}) is fulfilled, and the estimate (\ref{zeta}) of the displacement
at lowest order is reliable. As a result, the final  energy of the electrons initially
located at  $Z\!=\!0$ after the expulsion is  \cite{FioFedDeA13}
\be{}
H=m{}c^2\gamma_{e{\scriptscriptstyle M}},\qquad\qquad
\gamma_{e{\scriptscriptstyle M}}\simeq  1+2K \zeta^2.
\label{gammaeM}
\ee

The EM energy carried by the pulse within the pancake cylinder $V$ of radius $R$ is
\be
\E=\!\int_V\!\!dV\,\frac{\bE^{{\scriptscriptstyle\perp}2}\!\!+\!\bB^{{\scriptscriptstyle\perp}2}}{8\pi}
\!\simeq\!\frac{R^2}{4}\!\! \int_0^l\!\!\!\!d\xi\,\bE^{{\scriptscriptstyle\perp}2}
\!\simeq\!\frac{(\pi R)^2}{\lambda^2}\!\! \int_0^l\!\!\!\!d\xi\, \bA^{{\scriptscriptstyle\perp}2}
\!\simeq\!\frac{(mc^2\!\pi R)^2}{(e\lambda)^2}\!\! \int_0^l\!\!\!\!d\xi\,pw^2
\!=\!\frac{2(mc^2\!\pi R)^2}{(e\lambda)^2} Y_e^3(l)  \label{pulseEn}
\ee
Assuming for simplicity that the modulating amplitude $w$ is symmetric around $\xi_0\!=\!l/2$, 
it follows $Y_e^3(l)=2Y_e^3(\xi_0)$. Requiring $R\!=\!\nu\zeta$ (with some number \
$\nu\!\sim\!1$) \  we find by (\ref{zeta})
\be{}
\nu\zeta\!=\!R\!=\!\left[\!\frac{\nu\E (e\lambda)^2}{4(\pi m{}c^2)^2}\!\right]^{\frac 13}.            \label{blabla}
\ee
We consider two prototype modulating amplitudes with a unique maximum in $\xi_0=l/2$,
 and symmetric with respect to $\xi_0$, namely a gaussian and a fourth degree
polynomial with the same maximum point $\xi_0$, multiplied by the characteristic functions resp. of  $[0,l]$ and
of  $\left[\frac {l-l_p}2,\frac {l+l_p}2\right]$ \ (with $l_p\!\le\! l$ the length of the support of $w_p$): 
\bea
\ba{l}
w_g(\xi)\!=\!a_g\, \exp\!\left[-\frac{(\xi\!-\!\xi_0)^2}{2\sigma}\right]
\theta\!\left[\!\frac 14\!-\!\left(\!\frac\xi l\!-\!\frac12\!\right)^2 \right]\!,\qquad
w_p(\xi)\!=\!a_p\, \left[\!\frac 14\!-\!\left(\!\frac{\xi- \xi_0}{l_p}\right)^2\! \right]^2
\theta\!\left[\!\frac 14\!-\!\left(\!\frac{\xi- \xi_0}{l_p}\right)^2\! \right].
\ea              \label{wgwp}
\eea
Let $l'_g,l'_p$ be the widths at half height of  $w_g^2,w_p^2$.
We adjust the parameters $a_p,a_g,l_p,\sigma$  so that  $l'_g\!=\!l'_p$ and the pulses
corresponding to  $w_g,w_p$ have the same energy (\ref{pulseEn}),
i.e. $\int_{-\infty}^\infty\!d\xi\,w_p^2=\int_{-\infty}^\infty\!d\xi\,w_g^2$;  we  assume $l$ large enough (say $l\ge 3\sqrt{\sigma}$ ) for 
the integral of $w^2_g$ to be practically equal to its value for $l\!=\!\infty$.
By straightforward computations 
in the appendix we show that these two conditions amount to:
\be
2Y_e^3\!(\xi_0)= Y_e^3(l)= \frac p {1260}a_p^2l_p\simeq \frac p2 a_g^2\sqrt{\pi\sigma} , \qquad\qquad
0.4 l_p\!\simeq\!l'_p\!=\!l'_g\!=\!2\sqrt{\sigma\ln 2}.
\ee
Consequently, and by (\ref{pulseEn}-\ref{blabla}),
\be
Y_e^3\!(\xi_0)\!=\!\zeta\!=\!\left[\!\frac{\E (e\lambda)^2}{4(\nu\pi m{}c^2)^2}\!\right]^{\frac 13}
 \qquad \quad
\ba{ll}
\sigma\!=\!\frac{l'_g{}^2}{4\ln2}, \quad 
& a_g^2\!=\!\frac{4}{p \sqrt{\pi\sigma}}Y_e^3\!(\xi_0)
\!=\!\frac{8\sqrt{\ln 2}}{p \sqrt{\pi}l'_g}\left[\!\frac{\E (e\lambda)^2}
{4(\nu\pi m{}c^2)^2}\!\right]^{\frac 13},\\[10pt]
 l_p\!\simeq\!\frac{5l'_g}2, \quad   
& a_p^2 \!=\!\frac{1008}{pl_g' }Y_e^3\!(\xi_0)
\!=\!\frac{1008}{pl_g' } \left[\!\frac{\E (e\lambda)^2}{4(\nu\pi m{}c^2)^2}\!\right]^{\frac 13}.
\ea                                                       \label{varie}
\ee

The  laser machine at the Flame facility in Frascati can
shoot \cite{JovFedTanDeNGiz12,GizEtAl13} linearly polarized pulses \ ($p\!=\!1/2$) \ with  \ $\lambda\!\simeq\! 8\!\times\! 10^{-5}cm$, \
energy \ $\E\!=\!5 \!\times\! 10^{7}erg$, \  
an approximately gaussian longitudinal modulating amplitude
 with width at half height \ $l'\!\simeq\! 7.5\!\times\! 10^{-4}cm$,  \ and a radius $R$
which can be tuned by focalization in the range \ $10^{-4}\div 1\, cm$ . \
Imposing  $l'_g\!=\!l'_p\!=\!l'$, these data and  formulae  (\ref{pulseEn}-\ref{varie}) 
respectively yield for \ $\nu\!=\!1,2$
\bea
&& \nu\!=\!1:\qquad \quad
Y_e^3\!(\xi_0)\!=\!\zeta\!=\! 1.4 \!\times\! 10^{-3} cm,\qquad 
\ba{ll}
\displaystyle\sigma\!\simeq\!2\!\times\! 10^{-7} cm^2, \qquad 
&\displaystyle a_g  \!\simeq\!3.75,\\[10pt]
\displaystyle l_p\!\simeq\!1.875\!\times\! 10^{-3}cm, \qquad  
&\displaystyle a_p \!\simeq\!61.5.
\ea                                    \qquad \quad                 \label{Flame1}     \\[10pt]
&& \nu\!=\!2:\qquad \quad 
Y_e^3\!(\xi_0)\!=\!\zeta\!\simeq\! 8.8 \!\times\! 10^{-4} cm,\qquad 
\ba{ll}
\displaystyle\sigma\!\simeq\!2\!\times\! 10^{-7} cm^2, \qquad 
&\displaystyle a_g  \!\simeq\!2.98,\\[10pt]
\displaystyle l_p\!\simeq\!1.875\!\times\! 10^{-3}cm, \qquad   
&\displaystyle a_p \!\simeq\!48.8.
\ea                   \qquad  \quad                               \label{Flame2}
\eea
A plasma with $n_0\!\ge\! 10^{17}cm^{-3}$ is obtained by ionization
from an ultracold gas (typically, helium) jet in a vacuum chamber
hit by such an energetic laser pulse as soon as $\Gamma_i\!<\!1$. Here 
$\Gamma_i\!:=\!\sqrt{U_i/\kappa}\!=\!
\sqrt{2U_i/m{}c^2u_{e{\scriptscriptstyle}}^{{\scriptscriptstyle \perp}2}}\!\simeq\!
\sqrt{2U_i/m{}c^2pw^2(\xi)}$
($\kappa\!\equiv$kinetic energy) 
are the Keldysh parameters (the ionization potentials $U_i$ are about $24eV, \, 54eV$ 
for first and second ionization respectively). We define  the $l$ of 
(\ref{wgwp}) as the length of the $z$-interval where the Keldysh parameter of first ionization
corresponding to the gaussian pulse fulfills $\Gamma_i\le 1$; from
$1\!=\!\Gamma_i^2\!=\!4U_i/m{}c^2w_g^2(0)\!=\!4U_i/m{}c^2a_g^2\exp[-l^2/4\sigma]$ \ we find \ 
for either value of $\nu$ $l\!\simeq\!3\!\times\! 10^{-3}cm $, \  whence  $\xi_0\!\simeq\!1.5\!\times\! 10^{-3}cm $.
The ionization is practically complete and immediate \cite{Puk02,JovFedTanDeNGiz12}
 because the Keldysh parameter for double ionization reaches values $\Gamma_i\!<\!1/100$  very fast.

In the case $\nu\!=\!1$, choosing  e.g. \ $
\xi_1\!-\!\xi_0\!=\!l_p/20\simeq l'_g/8\simeq 0.06\xi_0$ \ 
and computing (\ref{Kxi1}) numerically (we have used the software {\it Mathematica}),
one finds \ $4 l'_g{}^2K_g(\xi_1)\!\simeq\! \xi_0^2K_g(\xi_1)\simeq 2$, \
$4 l'_p{}^2K_p(\xi_1)\!\simeq\! \xi_0^2K_p(\xi_1) \!\simeq\!2.2$, \
which by (\ref{budiffbound})corresponds to \ $n_0\!\simeq\!1\div1.1 \times 10^{18}cm^{-3}$.





\begin{figure}[ht]
\includegraphics[width=7.8cm]{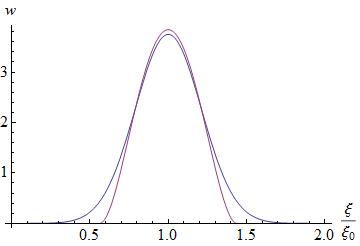}\hfill\includegraphics[width=7.8cm]{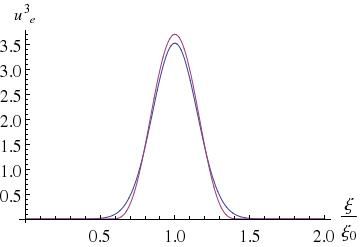}\\
\includegraphics[width=7.8cm]{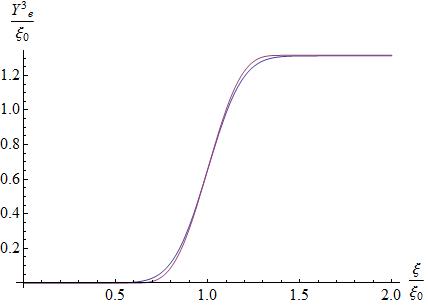}\hfill\includegraphics[width=7.8cm]{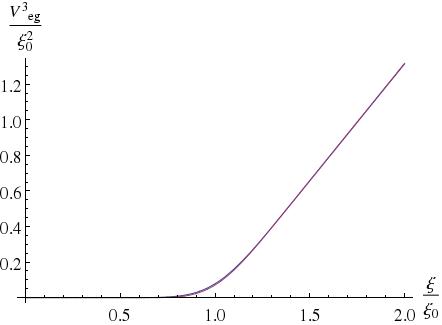}\\
 \includegraphics[width=7.8cm]{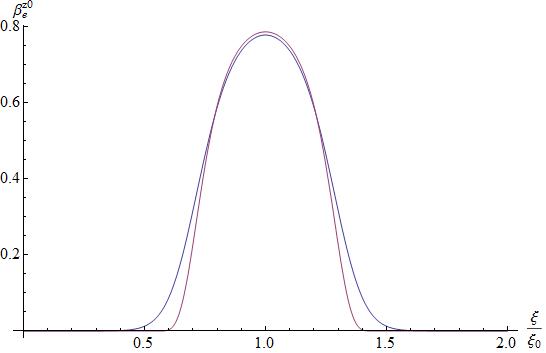}\hfill \includegraphics[width=7.8cm]{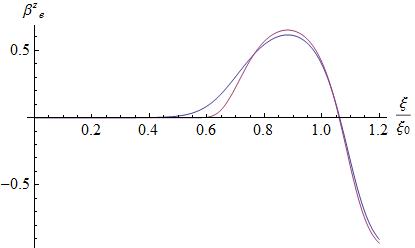}\\
\includegraphics[width=7.8cm]{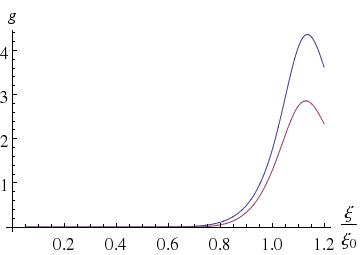}\hfill \includegraphics[width=7.8cm]{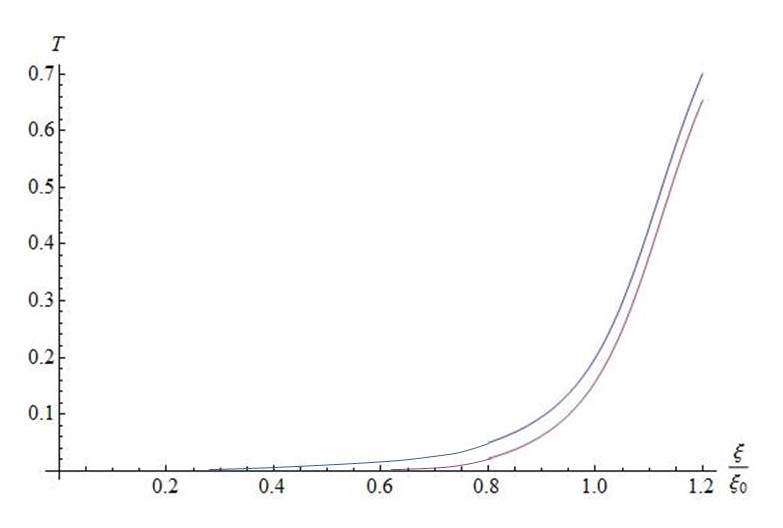}
\caption{The graphs of $w,u_e^{{\scriptscriptstyle z(0)}},Y^3_e,V^3_e,\beta_e^{{\scriptscriptstyle z(0)}},\beta_e^{{\scriptscriptstyle z(1)}},g,T$ for $w(\xi)\!=\!w_g(\xi)$ (blue curve) and 
 $w(\xi)\!=\!w_p(\xi)$ (purple curve) 
and $K\xi_0^2\!=\!2$, corresponding
to $n_0\!\simeq\! 10^{18}cm^{-3}$. \ In either case $T(\xi)$ is manifestly strictly increasing,
 and $T(\xi_0)\!\ll\! 1$.}
\label{Plots}       
\end{figure}

In  fig.  \ref{Plots} we plot $w_g(\xi)$ (blue line) and  $w_p(\xi)$ (purple line) together with the corresponding 
$u_e^{{\scriptscriptstyle z(0)}},Y^3_e,V^3_e,\beta_e^{{\scriptscriptstyle z(1)}},g,G,T$ for  circular
 polarization ($p \!=\!1$) and $\nu\!=\!1$, $a_g=3.75$, $a_p=61.5$, 
$2\xi_0\!=\!l\!=\!3\!\times\! 10^{-3}cm $, $K\xi_0^2\!=\!2$.
As manifest from these plots, the polynomial and the gaussian modulating amplitude
with the same energy and width at half height do not lead to significant differences.
As one can see from comparison of the graphs, the tiny tail of the gaussian 
outside the support of $w_p$ has a not completely negligible
effects, because it leads to a relative error  $T_g(\xi_0)\!\simeq\! 0.2$ instead of 
 $T_p(\xi_0)\!\simeq\!0.16$. However, as both $T_g(\xi_0),T_p(\xi_0)\!\ll\! 1$
we can consider (\ref{condgood}) with such a $K$ fulfilled and 
the above calculations leading to (\ref{zeta}) reliable. 
By (\ref{gammaeM}), this leads
to the slingshot effect with expulsion of the surface electrons with a  final  energy  $H\!\simeq\!2.3\, MeV$.

In the case $\nu\!=\!2$, choosing  \ $
\xi_1\!-\!\xi_0\!\simeq\! 0.19 l'_g\simeq 0.094\xi_0$ \ 
and computing (\ref{Kxi1}) numerically,
one finds \ $l'_g{}^2K_g(\xi_1)\!\simeq\! 0.32$, \
which by (\ref{budiffbound}) corresponds to \ $n_0\!\simeq\! 7 \times 10^{17}cm^{-3}$.
This leads to relative errors  $T_g(\xi_0)\!\simeq\! 0.22$,
 $T_p(\xi_0)\!\simeq\!0.18$. Again, as both $T_g(\xi_0),T_p(\xi_0)\!\ll\! 1$
we can consider (\ref{condgood}) with such a $K$ fulfilled and 
the above calculations leading to (\ref{zeta}) reliable. 
By (\ref{gammaeM}), this leads
to the slingshot effect with expulsion of the surface electrons with a  final  energy  $H\!\simeq\! 0.96\, MeV$.

\bigskip
{{\bf Acknowledgments.}} \ \ It is a pleasure to thank R. Fedele and U. De Angelis
for stimulating discussions and constant encouragement. I also acknowledge useful suggestions
on the use of {\it Mathematica} by S. De Nicola.

\section{Appendix}
\label{appendix}

{\bf 5.1 \ Proof of  (\ref{tilder}).} \ \ We use (\ref{defr}), (\ref{hatxtxp}), the inequality \
$z_e^{{\scriptscriptstyle(0)}}(x^0\!\!,\!Z)\!>\!0$ \ for all \ $x^0\!>\!Z$, \ 
change  integration variable  \
$s\!\mapsto\! y\!:=\!s\!-\!z_e^{{\scriptscriptstyle(0)}}(s\!,\!Z)
\!=\!\Xi_e^{-1}\!( s\!-\! Z)$ \ and abbreviate
 \ $\hat\xi(x^0\!,\! Z)\!:=\!\Xi_e^{-1}\!( x^0\!\!-\! Z)$:
$$
\ba{l}
\frac{\tilde r_e^{{\scriptscriptstyle(0)}}\!(x^0\!, Z )}{4K}\!=\!\!\int\limits^{x^0}_{Z }\!\!ds\frac {z_e^{{\scriptscriptstyle(0)}}\theta[z_e^{{\scriptscriptstyle(0)}}]\!-\! Z}{\tilde\gamma_e^{{\scriptscriptstyle(0)}}}(s\!,\! Z)\!=\!
\int\limits^{x^0}_Z\!\!ds \frac {z_e^{{\scriptscriptstyle(0)}}\!(s\!,Z)\!-\! Z}
{\gamma_e^{{\scriptscriptstyle(0)}}\![s\!-\!z_e^{{\scriptscriptstyle(0)}}\!(s\!,Z)]}
\!=\! \int\limits^{x^0}_Z\!\!ds\frac {s\!-\!Z \!-\!\Xi_e^{-1}\!(s\!-\! Z)}
{\gamma_e^{{\scriptscriptstyle(0)}}[\Xi_e^{-1}\!( s\!-\! Z)]}=
\!\!\int\limits^{\hat\xi}_0\!\! dy\,[\Xi_e\!(y)\!-\!y] \!=\!\!
\int\limits^{\hat\xi}_0\!\! dy\, Y^3_e\!(y)
=V_e^3\!\!\left[\hat\xi\right].
\ea
$$

\medskip
\noindent
{\bf  5.2 \ A  more explicit expression and a upper bound  for the rhs(\ref{appr1})$_1$.} \ \ \
Let $\underline{x}(\xim)$ be  the intersection (see fig.
\ref{trajectories4} -right) in the $(x^0, z)$ plane of the trajectory 
$(x^0, z_e^{{\scriptscriptstyle(0)}}(x^0\!,0))$ (plot in light orange) 
with the line of equation \ $x^0\!+\!z=\xim$, \ 
$\hat \xi(\xi,\xim')\!:=\!\min\{\xi, \underline{x^0}(\xim')\!-\!\underline{z}(\xim')\}$.
Using (\ref{lowestorderdensity}) we find \ $[n_e^{{\scriptscriptstyle(0)}}\bb^{{\scriptscriptstyle \perp(0)}}_e](x^0\!,z)\!=\!n_0\theta[Z_e^{{\scriptscriptstyle(0)}}(x^0\!,z)]\bu_e^{{\scriptscriptstyle(0)}}(x^0\!-\!z)$, whence
(using $d^2x'\!=\!\frac 12d\xim'd\xi'$)
\bea
\displaystyle\int^{\xi}_{ 0}\!\!\!\! d\xi'
[n_e^{{\scriptscriptstyle(0)}}\bb^{{\scriptscriptstyle \perp(0)}}_e](\xi'\!,\!\xim')
=\!\!\displaystyle\int^{\xi}_{ 0}\!\!\!\! d\xi'n_0\theta[Z_e^{{\scriptscriptstyle(0)}}\!(\xi'\!,\!\xim')]\bu_e^{{\scriptscriptstyle \perp(0)}}\!(\xi')=\!\!\!\!
\displaystyle\int\limits^{\hat\xi(\xi,\xim')}_{ 0}
\!\!\!\!\!\! d\xi' n_0\bu_e^{{\scriptscriptstyle \perp(0)}}\!(\xi')
\!=\!n_0\bY_e\!^{{\scriptscriptstyle \perp}}\!\!\left[\hat\xi\!\left(\xi\!,\!\xim'\right)\!\right]\!,
\quad \label{grillo}\\
 \displaystyle\int_{D\!_{x}\cap T}\!\!\!\!\!\!\!\!\!\! d^2x'[n_e^{{\scriptscriptstyle(0)}}\bb^{{\scriptscriptstyle \perp(0)}}_e](x')=\!\!
\displaystyle\int^{\xim}_{0}\!\!\!\!\!\!d\xim' \!\!
\displaystyle\int^{\xi}_{0}\!\!\!\!\!d\xi'[\frac{n_e^{{\scriptscriptstyle(0)}}}2\bb^{{\scriptscriptstyle \perp(0)}}_e](\xi'\!,\xim')=\frac {n_0}2\!\!
\displaystyle\int^{\xim}_{0}\!\!\!\!\!\!d\xim' \,\bY_e\!^{{\scriptscriptstyle \perp}}\!\!\left[\hat\xi\!\left(\xi,\xim'\right)\right]=:n_0\bW_e\!^{{\scriptscriptstyle \perp}}(x);\quad 
\eea
both integrals vanish for \ $\xi\!\le\! 0$, \ as so does
 $\bY_e\!^{{\scriptscriptstyle \perp}}\!\!\left(\xi\right)=0$. Thus (\ref{appr1})$_1$ becomes
\be
\bu_e^{{\scriptscriptstyle \perp(1)}} \!-\!
\bu_e^{{\scriptscriptstyle \perp(0)}}= \!-\!2K\,\bW_e\!^{{\scriptscriptstyle \perp}}
\qquad \qquad K\!=\!\frac{\pi e^2 n_0}{m_ec^2}   \label{budiff}
\ee
Consequently, for \ $ 0\!\le\! \xi\!\le\! \xi_0$ and the  ``pump''
$\Be^{{\scriptscriptstyle \perp}}\!=\!\epsilon_s\Be_o^{{\scriptscriptstyle \perp}}$
 with either polarization 
we find (\ref{budiffbound}):
$$
\frac{|\bu_e^{{\scriptscriptstyle \perp(1)}} \!-\!\bu_e^{{\scriptscriptstyle \perp(0)}}|}{w}\le 
 \!\displaystyle\int^{\xim}_{0}\!\!\!\!\!\!d\xim' \frac K{w(\xi)}\left|\bY_e\!^{{\scriptscriptstyle \perp}}\!\!\left[\hat\xi\!\left(\xi,\xim'\right)\right]\right|\le\!\!
\displaystyle\int^{\xim}_{0}\!\!\!\!\!\!d\xim' \frac {Kw\!\left[\hat\xi\!\left(\xi,\xim'\right)\right]}{kw(\xi)}\le \frac{K\lambda\xim}{2\pi };   
$$
the last inequality holds because $w(\xi)$ is increasing in $]0,\xi_0[$, and $\hat\xi\!\le\!\xi$.
The inequality is clearly valid also with the equal expression \ \
$k|\bA_e^{{\scriptscriptstyle \perp(1)}}(\xi,\xim)\!-\!
\Ba_e^{{\scriptscriptstyle \perp}}(\xi)|/\epsilon_s(\xi)$ \ at the lhs.

\bigskip
\noindent
{\bf 5.3 \  $Y_e^3,V_e^3$ in closed form in the case $w=w_p$.} \ \ 
With the shift \ $\tilde\xi:=\xi\!+\!(l_p\!-\!l)/2$ \ the definition  (\ref{wgwp})$_1$ takes the form
$w_p(\xi)\!=\!\tilde w_p(\tilde \xi)\!:=\!a_p\, \left[\!\frac 14\!-\!\left(\!\frac{\tilde\xi}{l_p}\!-\!\frac 12\right)^2\! \right]^2
\theta\!\left[\!\frac 14\!-\!\left(\!\frac{\tilde\xi}{l_p}\!-\!\frac 12\right)^2\! \right]$.
By some computation one finds
\be
\ba{l}
\frac {2}{p l_pa_p^2}\tilde Y_e^3(\tilde\xi)=\displaystyle\int^{\tilde\xi/l_p}_0\!\! dy\, y^4(y\!-\!l_p)^4=
\frac {1}9\frac{\tilde\xi^9}{l_p^9}-\frac {1}{2}\frac{\tilde\xi^8}{l_p^8}+\frac {6 }{7}\frac{\tilde\xi^7}{l_p^7}-\frac {2}{3}\frac{\tilde\xi^6}{l_p^6} +\frac {1}5\frac{\tilde\xi^5}{l_p^5},\\[12pt]
\frac {2}{pl_p^2 a_p^2}\tilde V_e^3(\tilde\xi)=\displaystyle\int^{\tilde\xi/l_p}_0\!\! dy\,\frac {2}{pl_pa_p^2}Y_e^3(y)=
\frac 1{90}\frac{\tilde\xi^{10}}{l_p^{10}}-\frac {1}{18}\frac{\tilde\xi^9}{l_p^9}+\frac {3 }{28}\frac{\tilde\xi^8}{l_p^8}
-\frac {2}{21}\frac{\tilde\xi^7}{l_p^7} +\frac {1}{30}\frac{\tilde\xi^6}{l_p^6}, 
\ea                                           \label{YV}
\ee
what gives at the maximum point of $\tilde w^2$
\ $\tilde Y_e^3(l_p/2)\!=\!p a_p^2l_p/2520$ \ 
 and  \ $\tilde V_e^3(l_p/2) \!=\!l_p \tilde Y_e^3(l_p/2) 63/512 
$,   \ 


\begin{thebibliography}{99}

\bibitem{FioFedDeA13} G. Fiore,  R. Fedele, U. De Angelis,
{\it The slingshot effect: a possible new laser-driven high energy acceleration
mechanism for electrons},  arXiv:1309.1400.

\bibitem{LanLif62}
L.D. Landau, E.M. Lifshitz, {\it The Classical Theory of Fields}, $2^{nd}$ ed. 
(translated from the Russian), Pergamon Press, 1962, p. 128.

\bibitem{EbeSle68}
J. H. Eberly,  A. Sleeper,
Phys. Rev. {\bf  176} (1968), 1570.

\bibitem{EbeSle69}
J. H. Eberly, 
Progress in Optics VII. (Ed. E. Wolf)  North-Holland, Amsterdam, 1969, pp 359-415;
and references therein.

\bibitem{Fioetal} G. Fiore, et al.,
{\it On a recursive determination of plane waves in relativistic cold plasmas}, 
in preparation.

\bibitem{Kar74} V. I. Karpman,  {\it Non-linear waves in dispersive media},
Pergamon Press, 1974.

\bibitem{She84} Y. R. Shen, {\it The principles of nonlinear optics},
New York, Wiley-Interscience, 1984.

\bibitem{Whi91} G. B. Whitham,  {\it Linear and Nonlinear Waves}, 
John Wiley \& Sons Inc., 1974.



\bibitem{Hec02}
E. Hecht,  {\it Optics}, Addison-Wesley, 2002, p.  67.

\bibitem{AkhEtAl75}
A.I. Akhiezer, R. V. Polovin, A. G.  Sitenko, K. N. Stepanov, 
 {\it Plasma electrodynamics, Vol. 1 - Linear theory}
Pergamon Press,  1975, p. 174.

\bibitem{TajDaw79}
T. Tajima, J. M. Dawson,
Phys. Rev. Lett. {\bf 43}, 267–270  (1979).

\bibitem{Gorbunov-Kirsanov1987} L.M. Gorbunov, and  V.I. Kirsanov, Sov. Phys. JETP \textbf{66}, 290 (1987).

\bibitem{Sprangle1988} P. Sprangle, E. Esarey, A. Ting, and G. Joyce, Appl. Phys. Lett. \textbf{53}, 2146 (1988).

\bibitem{Irman2007} A. Irman, M. J. H. Luttikhof, A. G. Khachatryan, F. A. van Goor, J. W. J. Verschuur, H. M. J. Bastiaens, K.-J. Boller, J. Appl. Phys. \textbf{102}, 024513 (2007).

\bibitem{Joshi2006} C. Joshi, 
Scientific American \textbf{294}, 40 (2006).

\bibitem{Fio14} G. Fiore, 
{\it On plane waves in diluted relativistic cold plasmas}, arXiv:1405.0163, to appear in Acta Appl. Math.

\bibitem{FauEtAl04}
J., Faure, Y. Glinec, A. Pukhov, S. Kiselev, S. Gordienko, E. Lefebvre, J.-P. Rousseau, F. Burgy, V. Malka, 
 Lett. Nat. {\bf 431}, 541–544 (2004).

\bibitem{MalEtAl05}
V. Malka, J. Faure, Y. Glinec, A. Pukhov,  J.-P. Rousseau,
 Phys. Plasmas {\bf 12} (2005), 056702.

\bibitem{Kalmykov2009} S. Kalmykov, S. A. Yi, V. Khudik, and G. Shvets, \emph{PRL} \textbf{103}, 135004 (2009).

\bibitem{AkhPol56}
A.I. Akhiezer, R.V. Polovin, 
(transl. from the Russian) Sov. Phys. JETP{\bf 3} (1956), 696.

\bibitem{KawDaw70} P. Kaw, J. Dawson,
Phys. Fluids 13 (1970), 472.

\bibitem{MouUms92} G. Mourou, D. Umstadter,  
    Phys. Fluids {\bf B4} (1992), 2315.

\bibitem{Daw59} J. D. Dawson, 
Phys. Rev. {\bf  113} (1959), 383.

\bibitem{AkhPol67} A. I. Akhiezer,  R. V. Polovin,
\textit{Collective oscillations in a plasma}, 
M.I.T. Press, 1967.

\bibitem{Puk02}
A. Pukhov, 
Rep. Prog. Phys. {\bf  65} (2002), R1-R55.

\bibitem{JovFedTanDeNGiz12}
D. Jovanovi\'c, R. Fedele, F. Tanjia, S. De Nicola,  L.A. Gizzi,
Eur. Phys. J. {\bf D66} (2012), 328.


\bibitem{GizEtAl13}
L.A. Gizzi, C. Benedetti, C.A. Cecchetti, G. Di Pirro, A. Gamucci, G. Gatti, A. Giulietti, D. Giulietti, P. Koester, L. Labate, T. Levatoy, N. Pathak, F. Piastra,  
Appl. Sci. 2013, {\bf 3}(3), 559-580.

\end{thebibliography}
\end{document}